# Self-Assembly of Rhamnolipids Bioamphiphiles: Understanding Structure-Properties Relationship using Small-Angle X-Ray Scattering

Niki Baccile,[a,*] Alexandre Poirier,[a] Javier Perez,[b] Petra Pernot,[c] Daniel Hermida-Merino,[d,e] Patrick Le Griel,[a] Christian C. Blesken,[f] Conrad Müller,[f] Lars M. Blank,[f] Till Tiso[f]

[a] Sorbonne Université, Centre National de la Recherche Scientifique, Laboratoire de Chimie de la Matière Condensée de Paris, LCMCP, F-75005 Paris, France

[b] Synchrotron Soleil, L'Orme des Merisiers, Saint-Aubin, Gif-sur-Yvette, France

[c] ESRF – The European Synchrotron, CS40220, 38043 Grenoble, France

[d] Netherlands Organisation for Scientific Research (NWO), DUBBLE@ESRF BP CS40220, Grenoble, 38043, France

[e] Departamento de Física Aplicada, CINBIO, Universidade de Vigo, Campus Lagoas-Marcosende, Vigo, 36310, Spain

[f] iAMB – Institute ofApplied Microbiology, ABBt – Aachen Biology and Biotechnology, RWTH Aachen University, Aachen, German

* Corresponding author:
Dr. Niki Baccile
E-mail address: niki.baccile@sorbonne-universite.fr
Phone: +33 1 44 27 56 77

**Abstract**

The structure-properties relationship of rhamnolipids, RLs, well known microbial bioamphiphiles (biosurfactants), is exlored in detail by coupling cryogenic transmission electron microscopy (cryo-TEM) and both *ex situ* and *in situ* small angle X-ray scattering (SAXS). The self-assembly of three RLs with reasoned variation of their molecular structure (RhaC10, RhaC10C10 and RhaRhaC10C10) and a rhamnose-free C10C10 fatty acid is studied in water as a function of pH. It is found that RhaC10 and RhaRhaC10C10 form micelles in a broad pH range and RhaC10C10 undergoes a micelle-to-vesicle transition from basic to acid pH occurring at pH 6.5. Modelling coupled to fitting SAXS data allows a good estimation of the hydrophobic core radius (or length), the hydrophilic shell thickness, the aggregation number and the surface area per RL. The essentially micellar morphology found for RhaC10 and RhaRhaC10C10, as well as the micelle-to-vesicle transition found for RhaC10C10, are reasonably well explained by employing the packing parameter (PP) model, provided a good estimation of the surface area per RL. On the contrary, the PP model fails to explain the lamellar phase found for the protonated RhaRhaC10C10 at acidic pH. The lamellar phase can only be explained by values of the surface area per RL being counterintuitively small for a di-rhamnose group and folding of the C10C10 chain. These structural features are only possible for a change in the conformation of the di-rhamnose group between the alkaline and acidic pH.

## Introduction

Bioamphiphiles, also known in the literature as biosurfactants, are amphiphilic compounds derived from natural resources and obtained by plant extraction, enzymatic synthesis, or microbial fermentation. The latter, certainly one of the most important families, has been known for decades[1–4] and it contains both glycosylated (sophorolipids, rhamnolipids, mannosylerythritol lipids, etc.) and peptidic (surfactin) lipids. Due to their lower environmental impact, bioamphiphiles have been developed for decades to replace synthetic surfactants.[2–4] Microbial bioamphiphiles are indeed considered to be more biodegradable and less toxic than petrochemical surfactants, and therefore they find use in a number of applications in detergency, cosmetics, environmental science, or as antimicrobial compounds,[5–7] with a milder effect on protein denaturation.[8]

If an impressive effort has been dedicated in the past years to bring microbial biosurfactants to the market,[9–13] with an increasing interest from industry,[14,15] their effective use in real-life applications will certainly depend on the knowledge and understanding of their physicochemical properties in water under diluted and semi-diluted conditions (typically between 0.1 and 20 wt%), classically representing surfactant concentrations in commercial products. For this reason, study of the surface tension, critical micelle concentration (cmc), solubility but also, and in particular, self-assembly is needed.

The study of biosurfactants' self-assembly beyond the cmc is a relatively recent topic of research[16] and recently reviewed by us.[17] According to the present state of the art, most studies concern surfactin, acidic and lactonic C18:1 sophorolipids, mannosylerythritol lipids (MELs) and rhamnolipids (RLs). The self-assembly properties of rhamnolipids are known to be strongly affected by their state of charge since the work of Ishigami in 1987[16] and Champion in 1995:[18] acidic pH promotes the vesicle phase, while alkaline pH promotes the micellar phase. However, these and many other works, of which the self-assembly properties were just reviewed,[19] were performed on sources of RLs containing both the di-rhamnolipid (RhaRhaC10C10) and mono-rhamnolipid (RhaC10C10),[16,18,20–24] thus precluding full understanding of the structure-properties relationship. Studies concentrating on either mono- or di-RLs started with the work of Chen *et al.*,[25] who showed that in buffer at pH 9 RhaRhaC10C10 tends to form micelles, while RhaC10C10 tends to form vesicles. However, data generated afterwards seemed to show that vesicles are observed for both RhaC10C10 and RhaRhaC10C10 around pH 6.[25] Other works have shown that RhaRhaC10C10 forms lamellar structures at acidic pH (< 5) and micellar at physiological pH (7.4).[26] Despite the fact that the amount of experimental work on the self-assembly of RLs is still quite limited, the impact of the buffer's choice is not to be excluded, as reported by Eismin et al. in terms of critical aggregation concentration (cac) and surface area per RL for a number of RL solutions.[27] More recently, a numerical modelling approach proposed that the acidic form of RhaC10C10 assembled into vesicles while its deprotonated form assembles into micelles.[28,29] If the existing data seem to show a difference in the assembly behaviour between RhaC10C10 and RhaRhaC10C10 with obvious pH effects (in relationship to the deprotonation of the free carboxylic acid), a clear relationship between the molecular structure of RLs and the morphology of its aggregates in water are still not fully understood.

With the goal of contributing to better understand this issue, this work reports a pH-dependent self-assembly study of three RLs, the classical RhaRhaC10C10 and RhaC10C10, but also a less known form, RhaC10, containing a single rhamnose unit and C10 chain, and a rhamnose-free C10C10, as negative control. Variation in the number of rhamnose units and C10 chains helps better understanding the relationship between the molecular structure of RL and its self-assembly properties in water. In particular, this works aims at evaluating the contribution of the C10 volume and rhamnose surface area to the packing parameter (PP)[30,31] expected for RLs, with PP being a simple model correlating the length and volume of the aliphatic moiety with the equilibrium surface area of the headgroup in amphiphiles. Previous studies have tried to correlate the PP calculated for RLs to the morphologies of their aggregated state in water,[25,32] but a consensus has not been found, yet, and discrepancies exist, as outlined by Chen *et al*.[25]

By employing both cryogenic transmission electron microscopy (cryo-TEM) and small angle X-ray scattering (SAXS), both *ex situ* and *in situ*, this work explores in a continuous manner the micelle-to-vesicle transition of RhaC10C10 and it compares it to the structures obtained for RhaRhaC10C10, RhaC10 and C10C10 in their protonated and non-protonated state. Quantitative parameters like the hydrophobic core radius (or length), the hydrophilic shell thickness, the aggregation number, surface area per RL are determined. The PP is calculated from these experimental data and compared to the PP calculated on the basis of the molecular structure. It is shown that, provided a good estimation of the value of the surface area per rhamnose, predicted PP meets reasonably well the self-assembled morphology for all RLs, except for the protonated form of RhaRhaC10C10, of which the lamellar structure observed at acidic pH can only be explained by a conformational rearrangement of the RhaRha group and important tilt of the C10C10 chain.

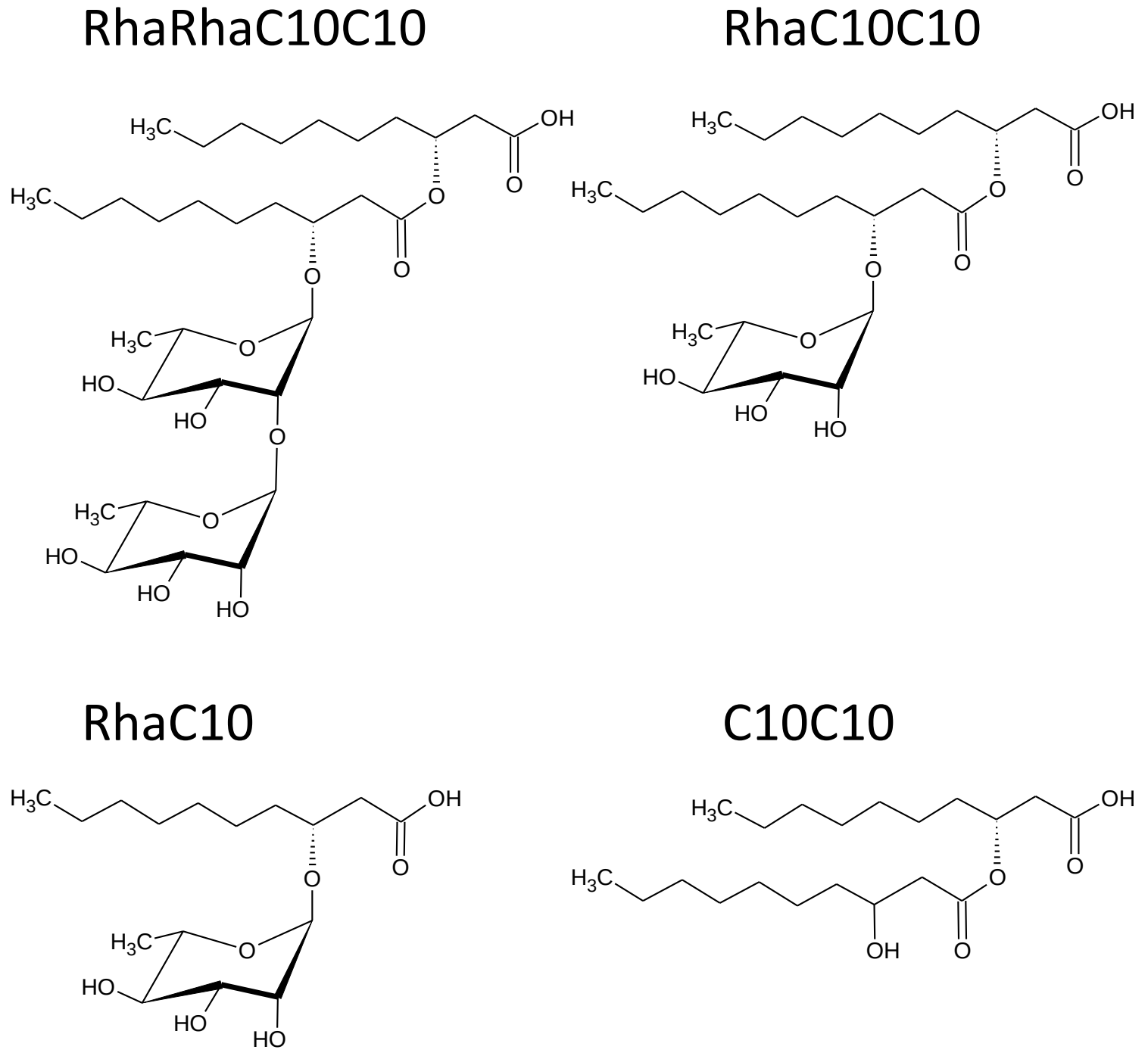

**Figure 1 – Chemical structures of the rhamnolipids studied in this work.**

## Material and Methods

*Chemicals.* Mono-rhamno mono-lipids (RhaC10) were purchased (Glycosurf, Salt Lake City, UT, USA). Mono-rhamno di-lipids (RhaC10C10), di-rhamno di-lipids (RhaRhaC10C10) and 3-(3-hydroxyalkanoyloxy)alkanoic acid (C10C10) were obtained by microbial production and subsequent purification.

For the production of C10C10, RhaC10C10 and RhaRhaC10C10, *P. putida* KT2440 KS3,[33] *P. putida* KT2440 SK4[34] and *P. putida* KT2440 pWJ02[35] were used, respectively. Rhamnolipid production was carried out in shake flasks using LB-medium (10 g/L tryptone, 5 g/L yeast extract, 10 g/L NaCl)[36] with additional 10 g/L glucose. The shake flasks were incubated at 30°C at 80 rpm (G25 Incubator shaker from New Brunswick Scientific Co. Inc. Enfield, USA, with a shaking diameter of 100 mm) in 1.8 L Fernbach flasks with 500 mL of culture volume for 3-4 days.

*Purification.* In-house purification was achieved by combining adsorption/desorption and semi-preparative liquid chromatography, as described before.[35,37] Briefly, proteins remaining in the fermentation broth after centrifugation were precipitated by solvent addition and removed via centrifugation. After evaporating the solvent, the supernatant was adsorbed using the silica adsorbent AA12SA5 (YMC Europe GmbH Dinslaken, Germany). Next, desorption of surfactants was carried out using ethanol, and the ethanol evaporated. For the chromatographic separation, a semi-preparative HPLC system (AZURA pump P6.1L, AZURA autosampler 3950 (both Knauer GmbH, Berlin, Germany), SEDEX 58 LT-ELSD detector (SEDERE Olivet, France), fraction collector Foxy R1 (Teledyne ISCO Lincoln, USA) equipped with a VP250/21 NUCLEODUR C18 HTec column (Macherey-Nagel GmbH & Co. KG, Düren, Germany) was employed. The flow rate was set to 10 mL/min, and the running buffers were acetonitrile and ultra-pure water with 0.2% (v/v) formic acid. After fractionation, the solvents were evaporated. The typical retention times with corresponding chromatograms for raw and purified samples are given in Table S 1.

*Quantification of surfactants*. RLs and C10C10 were quantified using high-performance liquid chromatography (HPLC) coupled with a corona-charged aerosol detector (CAD) with a method described previously[38] and based on methods developed earlier.[39,40] Biological samples were prepared as follows: The cell-free culture broth was mixed 1:1 with acetonitrile and stored overnight at 4 °C to precipitate the proteins. The samples were then centrifuged and filtered using Phenex RC syringe filters (0.2 µm, Ø 4 mm, Phenomenex, Torrance, USA). Next, reversed-phase chromatography coupled to CAD was performed to quantify rhamnolipid and C10C10 concentrations using an Ultimate 3000 with a corona Veo charged aerosol detector (Thermo Fisher Scientific, Waltham, MA, USA). For separation, a NUCLEODUR C18 Gravity 150 x 4.6 mm column (particle size: 3 mm, Macherey-Nagel GmbH & Co. KG, Düren, Germany) was used. The flow rate was set to 1 mL/min, and the column oven temperature was set to 40 °C. Acetonitrile and ultra-pure water with 0.2% (v/v) formic acid were used as running buffers.

*Determination of turbidity*. Turbidity of RLs and C10C10 solutions was determined via optical density at a wavelength of 860 nm using a Synergy MX plate reader (BioTek, Bad Friedrichshall, Germany). Measurements were performed in 96-well plates with a clear bottom (Greiner Bio-One, Kremsmünster, Austria), which were each filled with 250 µL of the test solutions immediately after sample preparation. Double distilled water was used for blank value calibration. If sample values exceeded 4.0, the corresponding samples were diluted by a factor of 2.

*Solubility assay*. To assess solubility C10C10 as a function of pH, starting from stock solutions with concentrations of 1.5 g/L and a pH of 7.5, samples were prepared with different pH values ranging from 1 to 13. Precipitation of C10C10 was evaluated by turbidimetry immediately after sample preparation. In addition, C10C10 concentrations were monitored over 14 days to investigate solubility in dependance of time.

*Sample preparation.* Samples are prepared by weighting the appropriate amount of RLs, to which a given volume of milliQ-grade water was added. Final concentrations were 5 mg/mL or 25 mg/mL. pH was adjusted using 0.1 M and 0.5 M HCl or NaOH solutions. The chosen concentrations are above the critical micelle concentration (cmc) of the RLs employed in this work both at acidic and basic pH (Figure S 1).

*pH measure.* pH is measured using a Hanna Scientific pH-meter, model HI 5221. The pH meter is connected to a computer, equipped with the fabricant's software [HI 92000, version 5.0.28], for automatic pH recording.

*Pendant drop tensiometry*. The drop shape analysis system DSA30 Krüss, Germany, is used with associated software and microsyringes SY20 of 1 mL in borosilicate glass. The cleanliness of the setup is verified by pumping 10 times the syringe volume with milliQ water. The surface tension must be constant and reproducible ± 0.5 mN/m during the total time of the experiment. A pendant drop of 11 – 30 mL of the solution is produced in air with a steel capillary having an external diameter of 1.83 mm. Images are recorded each 1 s during 300 s. Contour of the drop is fitted by the Young-Laplace equation using an iterative process with the surface tension, σ, as an adjustable parameter

*Small angle X-ray scattering (SAXS).* SAXS experiments were performed at room temperature (23°C) using synchrotron light on the following instruments, each having specific characteristics. Similar for all, $q$ is the wavevector, with $q = 4\pi/\lambda \sin(\theta)$, $2\theta$ corresponding to the scattering angle and $\lambda$ the wavelength. The $q$-range was generally calibrated between $\sim 0.05 < q$ / $nm^{-1} < \sim 5$. Raw data obtained on the 2D detector were integrated (after masking systematically wrong pixels and the beam stop shadow) azimuthally using the in-house software provided at the beamline and thus providing the typical scattered intensity $I(q)$ profile. Absolute intensity units are determined by measuring the scattering signal of water ($I_{(q=0)} = 0.0163\ cm^{-1}$). To avoid imperfections in the subtraction process, the background (water + capillary) was always recorded on the same capillary at the exact same spot of the analysis and it was subtracted to the integrated data. Calibration of the $q$-scale was done with silver behenate ($d_{(001)}$ = 58.38 Å) as standard.

1) SWING beamline (Synchrotron Soleil, Saint-Aubin, France) during the proposal N. 20201747. The energy of the beam was set at 12 KeV and sample-to-detector distance at 2.005 m. The experiments were performed in a pH-resolved mode. To do so, we used a flow-through quartz 1.5 mm capillary connected to the sample-containing solution at pH about 9.6 through a peristaltic pump. The pH was controlled *in situ* via a classical KCl pH-meter directly located in the experimental hutch and monitored in real time from the control room through a computer interface. We used a Hanna pH meter with its related acquisition software (described above). Acquisition was triggered manually with SAXS acquisition. Each pH and SAXS profile are recorded every 5 s. pH is added in a controlled manner using a push syringe controlled from the experimental hutch.

2) BM29 beamline (ESRF, Grenoble, France) during the proposal N. MX2311. The experiments were done at 12.5 KeV, with a sample-to-detector distance of 2.83 m. We used the standard automatic sample-changer environment available at BM29 beamline. We mostly used 96 well-plates as sample holder and injection volume was set at 100 µL for each sample.

3) BM26 beamline (ESRF, Grenoble, France) during the proposal N. SC-5125. The experiments were conducted at 12 KeV, with a sample-to-detector distance of 2.6 m. We used a home-made flow-through capillary to analyse the sample. The capillary was filled manually using a 1 mL syringe and rinsed afterwards with water and ethanol.

*Analysis of the SAXS data.* The pH-resolved *in-situ* SAXS data were analyzed using a model-dependent and model-independent approach. The low-$q$ region below $q<$ ~0.03 Å$^{-1}$ is analyzed using an absolute power law model, while the $q$-region above 0.03 Å$^{-1}$ is fitted with a model-dependent function, the general expression of which is $I(q) \propto P(q)\ S(q)$, where $P(q)$ is the form factor of the scattering object and $S(q)$ being the structure factor correlating objects in space. At infinite dilution and in the absence of forces, may them be attractive or repulsive, $S(q)$ approaches unity and $I(q)$ becomes proportional to $P(q)$ only. In the present work, $S(q)$ is unitary for most samples, except one specific case, discussed in detail later. All individual models are available as such in the SasView 3.1.2 software, which was used to fit the SAXS profiles. On the contrary, the Edit Custom Model functionality was employed to combine the models together so to fit the entire SAXS profile at once.

*Fit of SAXS data: use of the absolute power law function.* The intense scattering profile, generally observed below $q<\sim0.03$ Å$^{-1}$, is fitted using a standard power law function, shown in Eq. 1. The lack of a plateau at low-$q$ and the the exponent $\alpha$ generally indicate the existence of a smooth interface ($\alpha$= 4), mass ($\alpha$ between 2 and 3) or surface fractals ($\alpha$ between 3 and 4).[41]

$$I(q) = scale * q^{-\alpha} + background \qquad \text{Eq. 1}$$

The absolute power law function is characterized by three free parameters, the *scale*, the *background* and $\alpha$. The absolute power law function is combined with either the micelle model or the membrane model using the sum command in the Edit Custom Model functionality of the SasView 3.1.2 software.

*Fit of SAXS data: the ellipsoid micelle model (Em).* The general equation of the scatterer intensity for centrosymmetrical objects is (Eq. 2)

$$I(q) = \frac{scale}{V}\left(\rho - \rho_{solvent}\right)^2 P(q)\, S(q) + bkg \qquad \text{Eq. 2}$$

where, *scale* corresponds to the volume fraction, $V$ is the volume of the scatterer, $\rho$ is the Scattering Length Density (SLD) of the object, $\rho_{solvent}$ is the SLD of the solvent, $P(q)$ is the form factor of the object, *bkg* is a constant accounting for the background level and $S(q)$ is the structure factor, which is hypothesized as unity for most samples in the analyzed range of $q$-values.

Fitting of SAXS data with typical profiles of micellar structures can be performed with a number of different form factor models. Several ones were tested, from the most simple, typically a sphere with homogeneous density, to more complex ones. In principle, choice of the form factor model should always be kept at its simplest, where the number of free variables is reduced to as few as possible. However, in the present case, classical models, like the homogeneous or core-shell sphere form factors, do not satisfactorily match the experimental data, as they fail matching the high-$q$ portion of the SAXS profiles. Based on our previous experience in fitting SAXS profiles of microbial glycolipid amphiphiles,[42,43] this issue was solved by employing an ellipsoid of revolution characterized by core and shell regions with uneven thickness (CoreShellEllipsoidXT in the SasView 3.1.2 software[44]), schematized in Figure 2.

The analytical expression of the $P(q)$ for a core-shell ellipsoid of revolution model (*Em*) implemented in the software is provided in ref. [44]. In the *Em* model (Figure 2), $T_{shell}$ is the equatorial shell thickness, $L_{shell}$ is the polar shell thickness, $R_{core}$ is the equatorial core radius, $L_{core}$ is the polar core radius, $\rho_{core}$, $\rho_{shell}$, $\rho_{solvent}$ are the SLDs of the micellar core, shell and solvent (water), respectively. The model also defines $X_{core}= L_{core}/R_{core}$ and $X_{shell}= L_{shell}/T_{shell}$, the aspect ratio in the hydrophobic core and hydrophilic shell, respectively. The advantage of this model is that it can always be simplified by fixing specific values of the variables. For instance, according to Figure 2, when $X_{shell} = 1$ and $X_{core} = 1$, the model reduces to a more common core-shell sphere form factor.

If use of $X_{core}\neq 1$ is classical, as it defines anisotropic objects, the use of $X_{shell}$ is generally quite rare in modelling the form factor of micellar systems. Typically, $X_{shell}= 1$ identifies a homogeneous shell thickness, classical for surfactant micelles, while $X_{shell}\neq 1$ identifies an inhomogeneous shell, as supposed before for other glycolipid amphiphiles, like sophorolipids.[43] In that case, $X_{shell}\neq 1$ improved the fitting of the SAXS profiles, especially the first minimum of the form factor,[42,43] which could not be satisfactorily matched by modifying the polydispersity value. In this work, we have observed similar features: polydispersity does not improve the fitting process of the first minimum of the form factor oscillation, which is otherwise matched by using $X_{shell}\neq 1$. However, even if $X_{shell}\neq 1$ could indicate an uneven distribution of the molecule inside the micelle,[43] its physical interpretation must be approached with care and for this reason $X_{shell}$ was not treated as a truly free variable: in the main manuscript, we present data obtained with $X_{shell}= 1$ and in the Supporting Information we show the improvements on the fit when $X_{shell}= 0.5$ ans 0.1. Further comments about $X_{shell}$ will be given during discussion of the data.

The SLD was calculated using Eq. 3 through the SLD calculator tool available in the SasView 3.1.2 software and based on the formula:

$$\rho = \frac{\sum_{i}^{j} Z_i r_e}{v_M} \qquad \text{Eq. 3}$$

where $Z_i$ is the atomic number of the $i^{th}$ of $j$ atoms in a molecule of molecular volume $v_M$, $r_e$ is the classical electron radius or Thomson scattering length (2.8179 x $10^{-15}$ m), with $\rho_{solvent}$ being set to 9.4 x $10^{-6}$ Å$^{-2}$, a known value for water ($\rho_{H2O}$ in Figure 2). In principle, the SLD should be estimated for each rhamnolipid molecule from their density. In practice, rhamnolipids, like other microbial glycolipids, are characterized by two regions of different electron density, rhamnose in the headgroup and fatty acid in the tail. Such molecular structure justifies the use of two values for the SLD, one for the tail (core) and one for the headgroup (shell), whereas $\rho_{core}$ was set to 8.4 x $10^{-6}$ Å$^{-2}$, a typical value for a hydrocarbon chain.[43] $\rho_{core}$ is assumed to be constant to simplify the fitting process, although one should be aware of the fact that partial hydration of the core was seen before for surfactant micelles and, hence, should not be excluded.[45–47] In this case, $\rho_{core}$ will diverge from ideality, thus introducing a source of error in the fitting strategy. Concerning the carboxylic acid, covalently bonded to the tail, one can reasonably assume it to be at the frontier of the core ans shell, with a strong contribution to the latter, as experimentally found in this work.

The shell SLD, $\rho_{shell}$, is more complex to determine as it contains the contributions of the rhamnose moiety, water and counterions. A reasonable assumption was that $\rho_{shell}$ should be contained between the values of $H_2O$ and dehydrated carbohydrate, that is between about 9.4 and 14 x $10^{-6}$ Å$^{-2}$. A more reasonable estimation should consider values slightly above the solvent level, in the order of 10 x $10^{-6}$ Å$^{-2}$, and dehydrated rhamnose ($C_6H_{12}O_5$) which, employing a density of $d$= 1.4 g/cm$^3$, yields an SLD value of about 13 x $10^{-6}$ Å$^{-2}$. In terms of the contribution of the carboxylic acid, if one considers formic acid ($CH_2O_2$, $d$= 1.2 g/cm$^3$) as a model, its SLD can be estimated to 10.7 x $10^{-6}$ Å$^{-2}$. During the fitting process, even if $\rho_{shell}$ was initially set as a free parameter, the fit did not converge; for this reason, $\rho_{shell}$ was initially treated as a variable and eventually fixed, as it had no real impact on the fit. The overall quality of the fit is followed by the classical $\chi^2$ evolution test but the the actual error on the fitted parameters is estimated to be about ± 10%. Additional considerations on the fitting process are given in the *Fitting strategy* paragraph below.

Finally, the *Em* is combined with the *Power Law* model using the sum command in the Edit Custom Model functionality of the SasView 3.1.2 software. Table S 2 presents the full list of all fitting parameters. When they are fixed, the corresponding value is given.

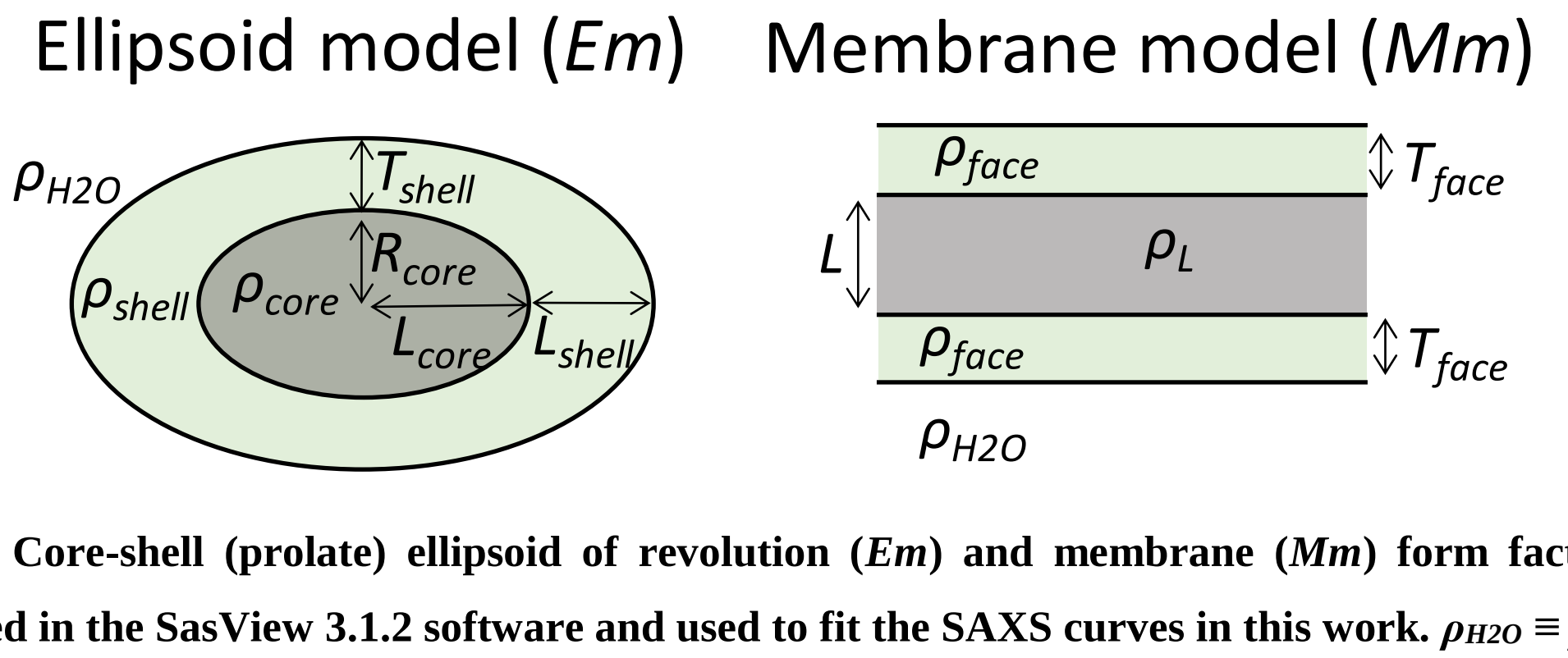

**Figure 2 – Core-shell (prolate) ellipsoid of revolution (*Em*) and membrane (*Mm*) form factor models implemented in the SasView 3.1.2 software and used to fit the SAXS curves in this work.** $\rho_{H2O} \equiv \rho_{solvent}$.

*Fit of SAXS data: the Membrane model (Mm).* The SAXS profiles typical of a bilayer membrane were fitted with a core-shell bicelle form factor (CoreShellBicelle in the SasView 3.1.2 software), which mimics a lipid bilayer membrane when the diameter of the bicelle is much larger than the bicelle's thickness. Here, the bicelle's radius parameter in the fit is set at the arbitrary value of 500 nm, with a thickness in the order of few nm, thus satisfying the condition. The model (geometrical sketch in Figure 2) is also characterized by a *scale* parameter, corresponding to the volume fraction, a background, a hydrophilic face region of given thickness, $T_{face}$, a hydrophobic core of given length, $L$, and a rim of given thickness, $T_{rim}$. Similarly to the micelle modell, the SLD of the face, core and solvent are $\rho_{core}$, $\rho_{face}$ and $\rho_{solvent}$ with $\rho_{core}$= 8.3 x $10^{-4}$ Å$^{-2}$ and $\rho_{solvent}$= 9.4 x $10^{-4}$ Å$^{-2}$. As for the *Em*, keeping $\rho_{face}$ as a free parameter did not allow the fit to converge. For this reason, $\rho_{face}$ was optimized manually although within the limits of 10.0 and 12.8 x $10^{-6}$ Å$^{-2}$, commented on above.

To reduce the number of independent parameters, the rim thickness was set to zero, as this parameter has no impact on the fit's quality. In this case, its corresponding SLD has no influence on the fit, either. The *Mm* model is also combined with the Absolute Power Law model and Table S 2 presents the full list of all fitting parameters.

*Fit of SAXS data: structure factor.* Introduction of a structure factor was necessary for one given system, of which the SAXS profile was characterized by a broad interaction peak. Considering the fact that rhamnolipids are charged, it was necessary to employ a typical structure factor based on screened coulomb repulsion between charged particles, implemented in the SasView 3.1.2 software as the Hayter rescaled mean spherical approximation (HayterMSA).[48,49] By setting the dielectric constant, salt concentration (here taken as negligible, as no salt is added), volume fraction and temperature, it is possible to estimate the micellar effective radius ($R_{Hayter}$) and *charge*. The HayterMSA structure factor is combined (multiplied, *S(q)*, Eq. 2) to the *Em* and *Power Law* (summed) models using the multiplication and summation tools command in the Edit Custom Model functionality of the SasView 3.1.2 software.

*Fitting strategy.* In this work, the full-scale fitting strategy was adopted, meaning that a single model combines more sub-models in order to fit the SAXS profile in the entire *q*-range. With the exception of one specific sample, for which a structure factor was needed, all samples are employed either by combination of *Em* + *Power Law* or *Mm* + *Power Law*. *Em* and *Mm* are used to fit the SAXS profiles at $q$> ~0.03 Å$^{-1}$, while *Power Law* is used to fit the SAXS curves at $q$> ~0.03 Å$^{-1}$, as illustrated in Figure S 2.

When combined, models always imply a large number of free variables. To reduce them, a typical SAXS profile is initially fitted using single models, *Power Law*, *Em* or *Mm*. The results of the fit are then employed as starting parameters in the combined model. However, each individual model is also characterized by a large number of variables. For this reason, as many variables as possible are calculated and fixed. Typically, *scale* (volume fraction), $\rho_{core}$ and $\rho_{solvent}$ are always fixed for both *Em* and *Mm*. Other parameters, like $\rho_{shell}$, $\rho_{face}$ or $X_{shell}$, did not allow the fit to converge, probably because their influence on the fit was limited, and for this reason, they were manually adjusted within the physical ranges discussed above so to find the best suited value. These parameters are eventually fixed for the rest of the fitting process. By doing so, the number of free variables becomes reasonably restrained to about three to four. To limit divergence of the fit, free variables are first fitted individually (e.g., $R_{core}$, $T_{shell}$, or $X_{core}$), followed by pairs (e.g., $R_{core}$ and $T_{shell}$ or $R_{core}$ and $X_{core}$), followed by triplets ($R_{core}$, $T_{shell}$ and $X_{core}$ at the same time). This is repeated for different variables alternatively. Eventually, when the most physically-reasonable values are obtained, all variables are fitted at the same time. If the fit converges and the values are simply refined, these are taken as good and reported in Table S 2. If the fit diverges, the whole process is repeated several times until the fit converges. However, in some cases, the fit always diverges. In this case, the parameters having the least influence on the fit (e.g., $X_{shell}$ or $\rho_{shell}$) are fixed. Once the fit converges, the values are considered as good and reported in Table S 2. The uncertainty of this process was estimated to ±10%. This value may seem high, but it reasonably takes into account the uncertainty of fixed variables, like the actual volume fraction (*scale*) or $\rho_{core}$.

Finally, polydispersity, which smoothes out the oscillation minima of the form factor, was set to zero and justified as follows. Typically, polydispersity is generally in the order of 0.1 to 0.2 for micellar systems, but in the present case, we observed that it does not have any practical impact, as the minima of the form factor can be fitted by the combination of the selected model and the corresponding fitting parameters. In addition, in a multi-parametric system, polydispersity can be applied to most variables (e.g., $R_{core}$, $L_{shell}$, $X_{shell}$, etc.), thus adding more complexity to the overall fitting process. In practice, use of zero polydispersity in the model does not mean that the described systems are not polydisperse, but the large error (10%) of the entire fitting process indirectly includes polydispersity in the results.

*Cryogenic transmission electron microscopy (cryo-TEM).* Microscopy experiments are performed using several instruments according to the resolution required. Ultra high resolution is achieved using a cryogenic transmission electron microscope (cryo-TEM, FEI Tecnai 120 twin microscope operated at 120 kV and equipped with a Gatan Orius CCD numeric camera). Lower resolution is obtained with an optical microscope equipped with polarized light (transmission Zeiss AxioImager A2 POL optical microscope equipped with an AxioCam CCD camera). Concerning the sample preparation, a drop of sample solution was settled on a holey carbon coated TEM copper grid (Quantifoil R2/2, Germany) and after removal of excess solution on the grid, it was plunged into liquid ethane and then stored at -180°C in liquid nitrogen until observation on the microscope.

## Results and Discussion

The self-assembly properties of rhamnolipids were explored by a combination of SAXS and cryo-TEM on three purified congeners, RhaRhaC10C10 and RhaC10C10, the classical components of a raw RL mixture, and RhaC10, produced synthetically. As a control, the same study was conducted on the rhamnose-free C10C10 compound. Except for the commercial one, all compounds were isolated from raw mixtures using chromatography (Table S 1) following previous procedures.[35,37]

Figure 3 shows the SAXS profiles of all RLs and C10C10 dispersed in water. All SAXS profiles were fitted using appropriate form factor models (see *Materials and Methods* section for more information on fitting), chosen on the basis of the typical shape of the SAXS curve but also on previous data published in the literature on both RLs[21,25] and other similar glycolipid bioamphiphiles.[42,43] The pKa of RLs is below 6, reported to be 5.9 for RhaC10C10 and 5.6 for RhaRhaC10C10 by Ishigami[16] and globally between 4.3 and 5.5 depending on concentration.[50] In this regard, and in analogy to the pH-dependent self-assembly properties of other microbial glycolipids, one expects phase transitions above pH 6, with little difference between neutral and strongly basic pH. For this reason, the SAXS profiles in Figure 3 were recorded at pH values spanning from slightly acidic to neutral. This choice was good enough to observe significant evolutions in the micellar characteristics and structures.

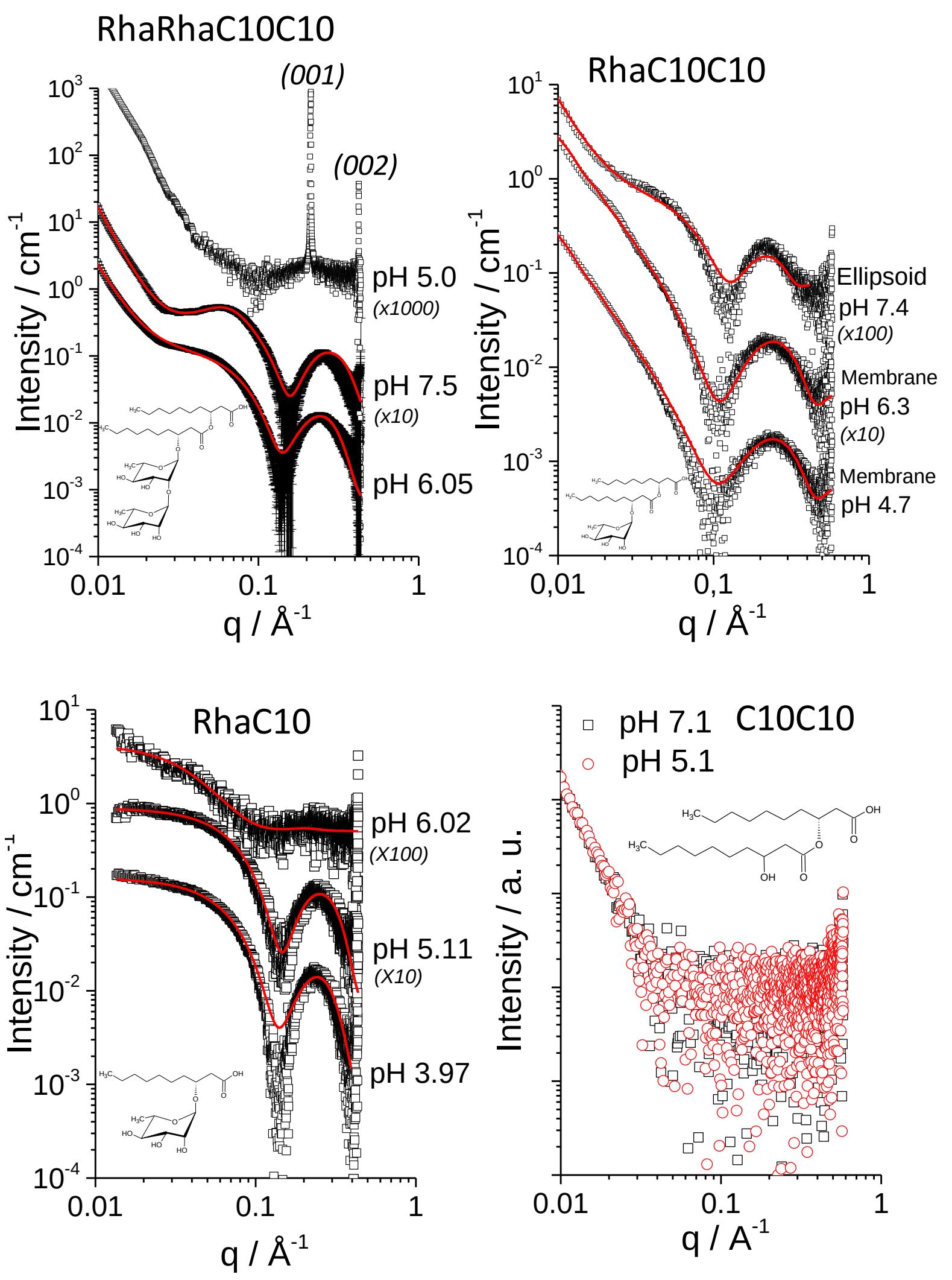

**Figure 3 – SAXS profiles (in absolute intensity, multiplication factors given next to each experiment) of rhamnolipid solutions. Concentrations are 25 mg/mL (RhaRhaC10C10, RhaC10), 5 mg/mL (RhaC10C10, C10C10). The full list of fitting parameters are given in Table S 2.**

*Qualitative analysis of RL SAXS profiles*

A qualitative description of the SAXS profiles gives the following. The RhaRhaC10C10 above pH 6 (Figure 3) shows a micellar profile combined with a strong low-*q* scattering, generally indicative of coexisting large-scale structures, often undefined. These are not uncommon, they were reported before, using both x-Ray[51] and neutron[20,52] scattering (thus ruling out possible beam damage effects), for microbial glycolipids dispersed in water,[20,51–53] and in the best case scenario they were described as minor amounts of nanoscale platelets[51] or residual self-assembled structures of the low pH region.[42]

To be noted, at pH 7.5, above the pKa, a broad hump reflects the presence of repulsive intermolecular interactions of electrostatic origin, typical of charged micelles in water.[54,55] At more acidic pH, below the pKa of RLs, RhaRhaC10C10 precipitates as a lamellar powder. This conclusion is drawn by the combination of the typical sharp (001) and (002) reflections of the corresponding SAXS profile and the presence of flat morphologies, observed by cryo-TEM (Figure 4). The latter excludes the presence of multilamellar vesicles, as sometimes found for other RLs, like RhaC10C10.[16,18,56] Similar results were recently reported by Ortiz et al. for RhaRhaC10C10 at pH 4.5.[26] Authors reported the formation of lamellar peaks, although it was unclear whether these were attributable to multilamellar vesicles (MLV) or lamellar precipitates. The present cryo-TEM experiment (Figure 4) rules out the presence of MLV for RhaRhaC10C10 below the pKa of RLs.

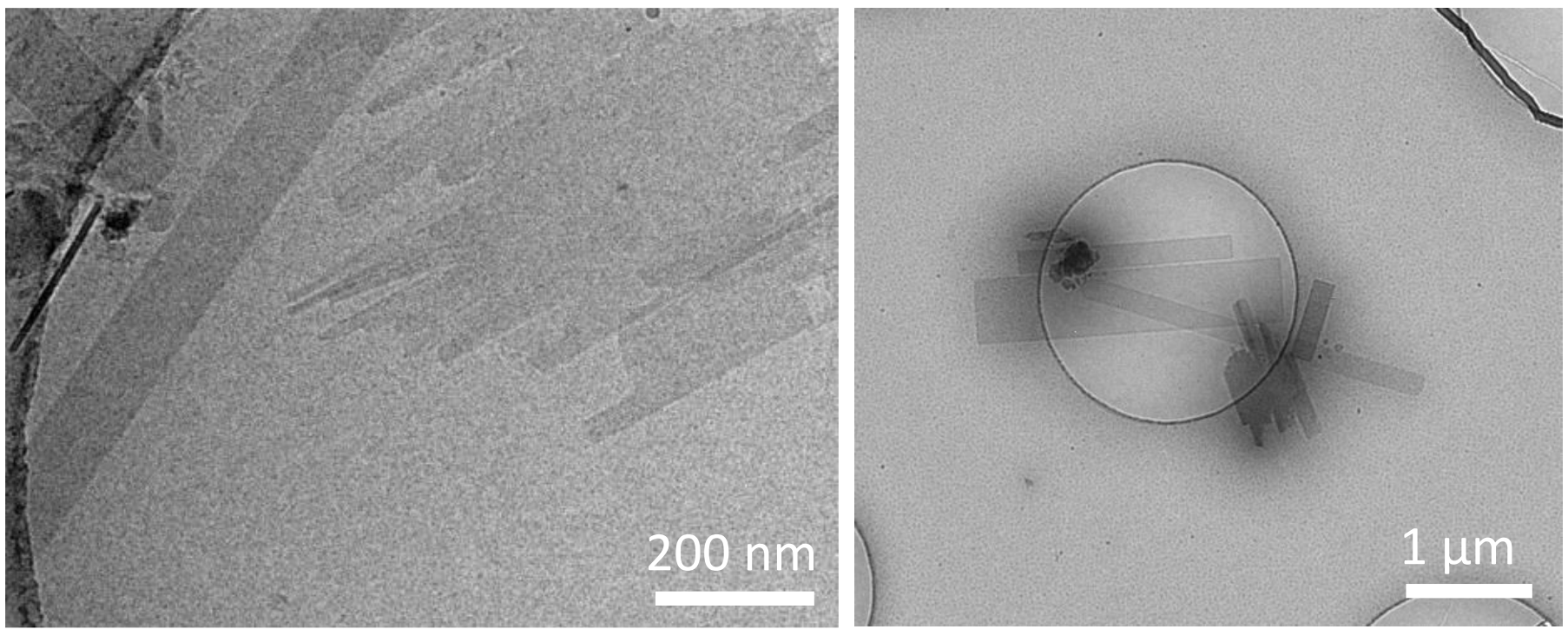

**Figure 4 – Cryo-TEM image of a RhaRhaC10C10 solution at pH 6.3 and 5 mg/mL.**

The RhaC10C10 (Figure 1) sample shows an evolution from a typical micellar profile at pH 7.4, to a signal displaying a broader oscillation of the form factor above 0.1 Å$^{-1}$ and a -2 dependence of the intensity against $q$ in log-log scale. The latter are typical characteristics of two dimensional membranes. This result is corroborated by cryo-TEM experiments (Figure 5a,b), showing the presence of unilamellar vesicles. Previous literature work shows that RhaC10C10 can form micelles at basic pH[27] and vesicles at acidic pH.[22,25,54] A minor content of vesicular aggregates were also reported at basic pH,[27] thus possibly explaining the strong low-$q$ scattering signal below 0.03 Å$^{-1}$ (Figure 3) for RhaC10C10 at pH 7.4. Interestingly, not only vesicles are not uncommon for other microbial glycolipids,[42,57] but also coexistence of membranes and micelles was observed in the micellar region of other microbial glucolipids.[57]

The RhaC10 (Figure 1) in the acidic pH regime presents SAXS patterns having similar profiles as both RhaRhaC10C10 at low pH and RhaC10C10 at neutral pH, except for the lack of the strong low-$q$ scattering signal. Interestingly, above pH 6, the signal undergoes a strong variation, with almost a complete loss in the oscillation of the form factor. Similar features were reported before for other microbial glycolipids under neutral-alkaline conditions[42,57] and are indicative of a strong hydration of the micelles, with possible rearrangement of the molecules inside the micellar aggregate.

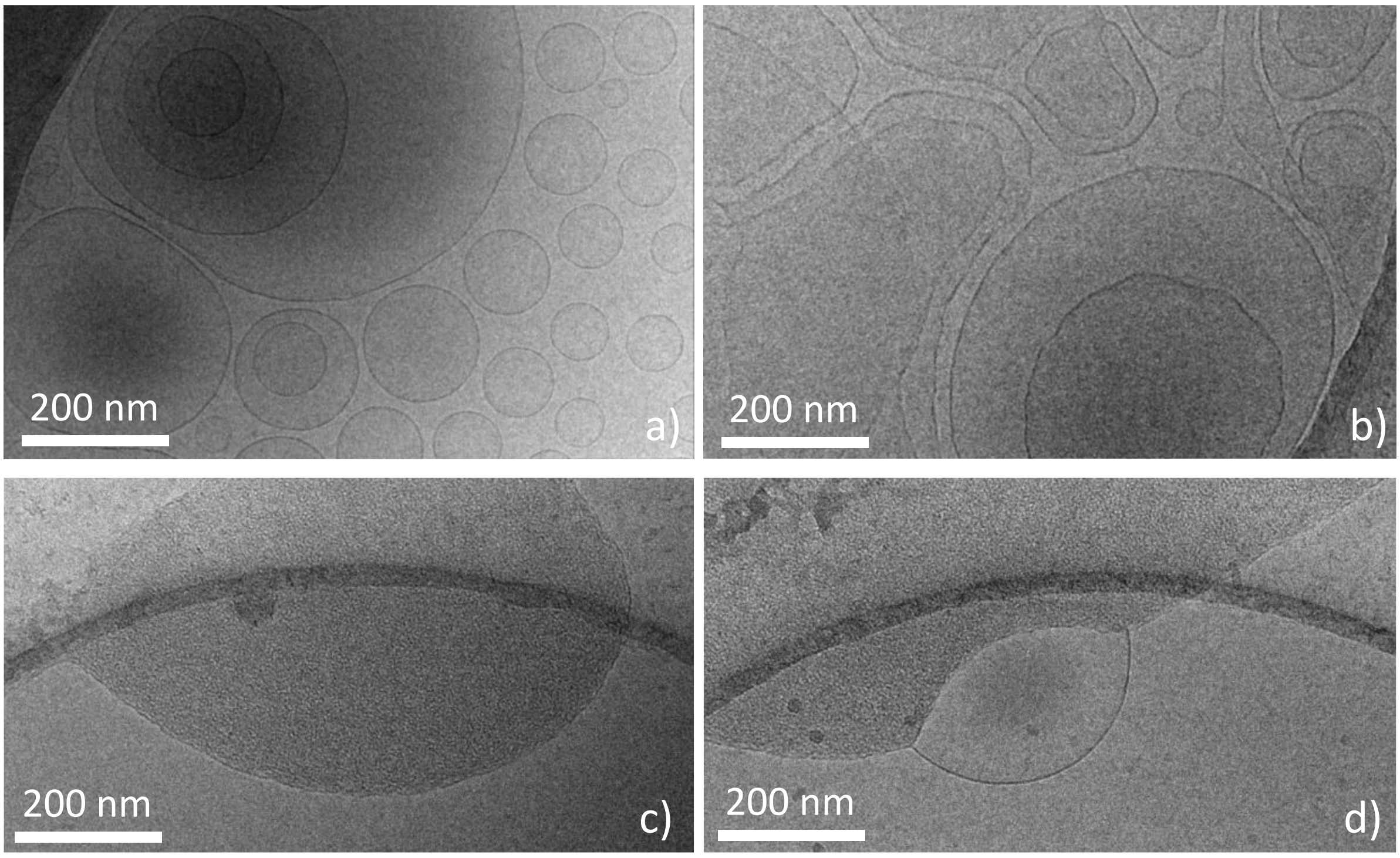

**Figure 5 – Cryo-TEM image of a RhaC10C10 solution (5 mg/mL) at a,b) pH 5 and c,d) pH 3.**

Finally, the rhamnose-free C10C10 (Figure 1) control shows a completely different behaviour from all other molecules. C10C10 is soluble under alkaline conditions (Figure S 3) but insoluble in the neutral-acidic pH range, as also confirmed by its SAXS patterns below pH 7.1 and being indicative of an insoluble precipitate. The corresponding cryo-TEM image of a C10C10 solution at pH 5 shows the presence of oil droplets in water, which are recognized by the dense homogeneous contrast (Figure 6). A dense droplet with a sharp interface is coherent with the strong low-$q$ scattering profile and no oscillation at high $q$, shown by SAXS. The absence of a dense corona at the outermost region of the droplets rules out the presence of vesicles, of which the typical cryo-TEM images are shown in Figure 5a,b.

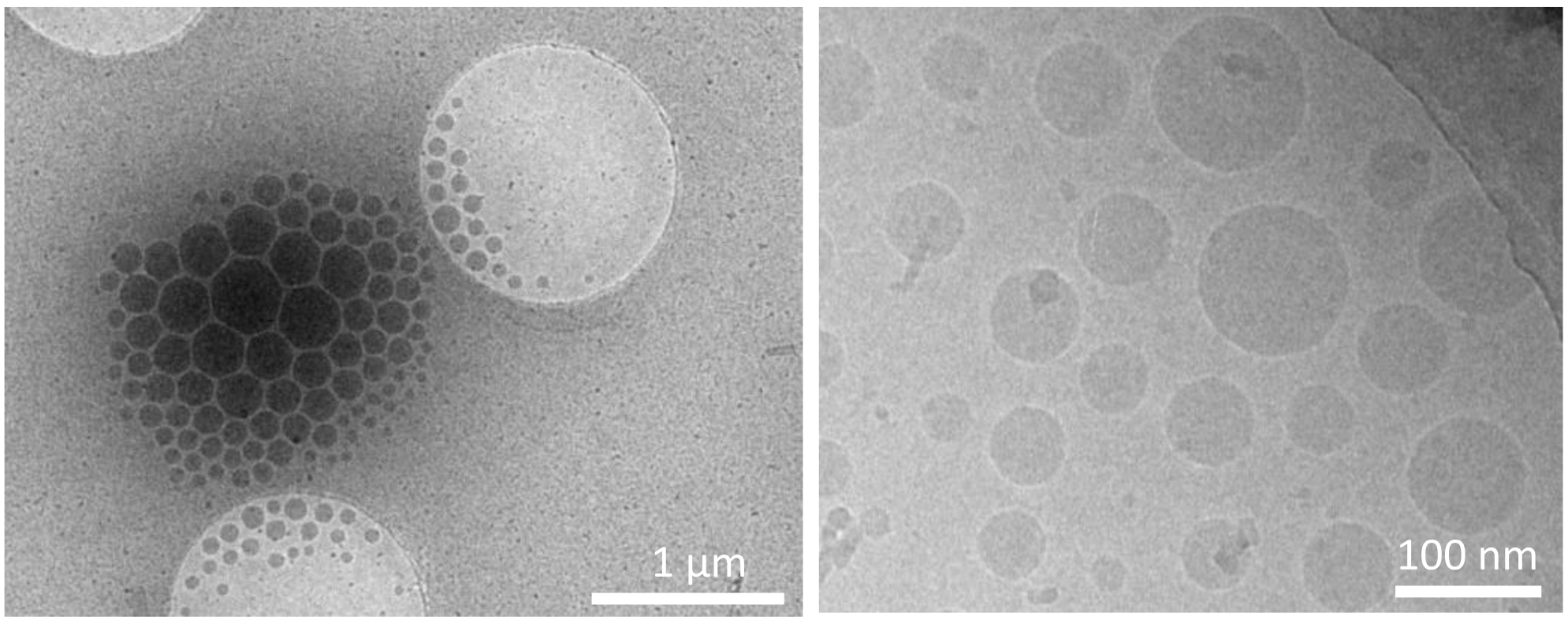

**Figure 6 – Cryo-TEM image of a C10C10 solution (5 mg/mL) at pH 5.**

*Quantitative analysis of RL SAXS profiles*

The quantitative analysis issued from the fitting process of the SAXS data is reported in Table 1. All micellar systems are fitted with an (prolate) ellipsoid of revolution form factor model function (the *Em* model), possibly coupled to a power law function for those systems presenting low-*q* scattering (Figure S 2). The *Em* model was applied before in modelling aqueous solutions of RLs,[21] but differently than in ref. [21], it is employed here as general as possible, similar to others works performed on microbial glycolipids.[42,43] This model, which describes the micelle as being constituted by a hydrophilic shell and a hydrophobic core, supposes that the shell thickness is not homogeneous. In the model, the thickness along the equatorial axis ($T_{shell}$, Figure 2) may be different than the thickness in the polar axis ($L_{shell}$, Figure 2). In terms of its physical meaning, $L_{shell} \neq T_{shell}$ means that the distribution of the rhamnolipids at the micelle-water interface is not homogenous,[43] as one would expect in typical head-tail surfactant micelles.[58] Non-micellar systems are modelled with a membrane model (*Mm*) with hydrophobic core and hydrophilic face. *Mm* is also coupled to a power law function in the low-*q* region (Figure S 2). More details on the fitting process are given in the *Materials & Methods*. The error associated to the fitting process is ±10%.

In the RhaRhaC10C10 at pH 6.05 and 7.5, micelles have a similar core radius of $R_{core}$= 8 ± 0.8 Å and a comparable shell thickness $T_{shell}$ in the order of 10 Å, both being in good agreement with previous estimations of both core and shell.[21] On the other hand, micelles are more elongated at pH 6.05 ($X_{core}$ = 10.8) than at pH 7.5 ($X_{core}$ = 2.5), at which a net charge of 6.4 appears, and attributed to the deprotonation of the carboxylic acid. The solution self-assembly behaviour of RhaRhaC10C10 is then quite straightforward. In the neutral form, RhaRhaC10C10 forms strongly anisotropic, neutral, micelles. The negative charges introduced upon deprotonation do not alter the cross section radius, but they introduce repulsive electrostatic forces, which increase the curvature of the elongated micelles into ellipsoids (smaller $X_{core}$) as well as long-range intermicellar interactions (broad hump at about $q$= 0.05 Å$^{-1}$). Parameters in Table 1 were obtained with a homogeneous shell thickness ($X_{shell}$= 1). However, the fit can actually be improved with non-unitary values of $X_{shell}$, typically $X_{shell}$= 0.1 and 0.5, as shown in Figure S 4, at both pH values. Off-unity values of $X_{shell}$ were reported before for other microbial glycolipid amphiphiles[42,43] and in each case, their meaning should interpreted with care: in the present case, $X_{shell} \neq 1$ could suggest an uneven distribution of the rhamnose and carboxylic acid/carboxylate (depending on pH) in the more hydrophilic region of the micelle.

At low pH, in the lamellar phase region, the peak position at $q$= 0.215 Å$^{-1}$ identifies an interlamellar distance of 29.2 Å, which includes the thickness of the membrane and the intermembrane water layer.

The RhaC10 molecule is micellar below and above the pKa of RLs, although with important structural changes. At acidic pH, below 5.11, below the pKa of RLs, micelles are essentially prolate ellipsoids ($X_{core}$ ~3) with equatorial core radius and shell thickness ($R_{core}$ of about 8 Å and $T_{shell}$ of about 10 Å, within the fit error) comparable to the neutral RhaRhaC10C10 micelles. Just as for RhaRhaC10C10, fits are of improved quality for $X_{shell}$< 1, as shown by the comparison presented in Figure S 5, where $X_{shell}$ varies between 0.1 and 1, possibly meaning that the equatorial direction is rich in the rhamnose group. Above the pKa, the scattering profile has completely changed, with strong similarities to the scattering profiles of other microbial glycolipids in their deprotonated form,[42] for which the loss in the oscillation of the form factor above 0.1 Å$^{-1}$ was associated either to a poor contrast between the hydrophilic shell and water, most likely associated to a strong hydration of the shell, or to a thin shell, in the order of 5 Å. Table 1 shows that the shell thickness decreases from 9.5 ± 0.9 at pH 5.11 to 3.4 ± 0.3 Å at pH 6.02, thus corroborating the data recorded on other microbial glycolipids. Elongation of the micelles is also possible ($X_{core}$= 9.1), although to be taken with care, considering the poor quality of the scattering profile. Visual observation combined with SAXS indicate that RhaC10 is water-soluble in a very large pH range, from acidic to basic. RhaC10 forms well-defined ellipsoidal micelles below the pKa of RLs. Above the pKa, the molecule seems to be more soluble with poor tendency to form micelles (noisy SAXS signal), which are strongly hydrated with poorly defined core-shell-water interface.

**Table 1 – Structural parameters derived from the fit of SAXS profiles provided in Figure 3. The models details are given in Figure 2 and the list of full parameters are given in Table S 2. In *Em**, a structure factor is employed. $V_{core}$ is calculated with Eq. 4, $N_{agg}$ is calculated with Eq. 8, with the volume of a single C10 chain taken as 243 Å$^3$ (Table S 5). Error on the fit parameters is ±10%. *: these data intervals are taken from the pH-resolved in situ experiment shown in Figure 7.**

| **Parameter** | **RhaRhaC10C10** | | | **RhaC10C10** | | **RhaC10** | | |
|---|---|---|---|---|---|---|---|---|
| pH | 5.0 | 6.05 | 7.5 | >7.5 | <6.5 | 3.97 | 5.11 | 6.02 |
| Phase | L | M | M | M | V | M | M | M |
| Model | - | *Em* | *Em** | *Em* | *Mm* | *Em* | *Em* | *Em* |
| $R_{core}$ / Å | - | 8.3 | 8.0 | 11.9 | - | 8.6 | 8.3 | 6.1 |
| $T_{shell}$ / Å | - | 10.8 | 8.2 | 7.7 | - | 10.6 | 9.5 | 3.4 |
| $X_{core}$ ($L_{core}/R_{core}$) | - | 3.1 | 2.5 | 2.0-2.9* | - | 3.1 | 2.5 | 9.1 |
| $V_{core}$ / Å$^3$ | | 7421 | 5359 | 17638 | | 8255 | 5985 | 8648 |
| $N_{agg}$ | | 15 | 11 | 36 | | 34 | 25 | 36 |
| $X_{shell}$ ($L_{shell}/T_{shell}$) | - | 1.0 | 1.0 | 1.0 | - | 1.0 | 1.1 | 1.1 |
| $R_{Hayter}$ / Å | - | - | 19.3 | - | - | - | - | - |
| *Charge* | - | - | 6.4 | - | - | - | - | - |
| $T_{face}$ / Å | - | | | | 5.5-6.1* | | | |
| *L* / Å | - | | | | 14.1-13.9* | | | |
| $d_{(2Tface+L+H2O)}$ / Å | 29.2 | | | | | | | |

The RhaC10C10 molecule, assembling into micelles above the pKa, displays a core radius 40% larger ($R_{core}$= 11.9 ± 1.2 Å) and an equatorial shell 29% smaller ($T_{shell}$ = 7.7 ± 0.8 Å) than the micelles composed of the other RLs studied in this work. Considering the error in the fit being 10%, these differences are small but significant and could be explained by the possible interpenetration between the chain and headgroup, as proposed by others.[25] In the acidic pH region, the unilamellar membrane of the vesicle phase displays a face thickness in the order of 6 Å and a core length of about 14 Å (Table 1) (refer to Figure 2 for the details of the ellipsoid and membrane models). To better understand the micelle-to-vesicle transition, and in particular the structural evolution between these two phases, well-known for RL mixtures and RhaC10C10 in particular,[20,22,26,27,59] a dedicated pH-resolved *in situ* SAXS study, shown in Figure 7, was attempted for the first time on RLs.

The pH-resolved *in situ* SAXS experiment is performed by adding a continuous flow of a HCl solution to a RhaC10C10 solution at pH 9, using the same setup employed for similar studies.[42,60] Figure 7 shows the experimental SAXS profiles recorded with pH, the *Em* and *Mm* models used to fit the SAXS curves and the pH-dependent evolution of the variables in the fit, associated to the following structural parameters: the core aspect ratio ($X_{core}$= $L_{core}/R_{core}$), the size of the core and shell (or face) regions ($R_{core}$, $T_{shell}$, $L$, $T_{face}$) and the slope, $\alpha$, corresponding to the exponent in the power law model, measured independently at $q$< 0.03 Å$^{-1}$ in the log-log scale. The evolution of $\chi^2$ helps visualizing the quality of the fit. In this regard, Figure S 6 shows a selection of fitted SAXS profiles of RhaC10C10 from acidic to basic pH using both the *Em* and *Mm* models. It can be seen that the *Em* model fits the experimental data reasonably well above pH 7 only, while the *Mm* model allows a reasonable fit of the databelow about pH 6.5. The evolution of $\chi^2$ in Figure 7 reflects very well this trend and it defines precisely the pH range of validity of each model. The shaded region in the plot of $\chi^2$ identifies the pH range in which both models fails. As a consequence, only the data reported in the non-shaded region are related to good quality fits and are meaningful within the context of this study.

The pH-dependent evolution of $\alpha$ shows precisely the pH range of the phase transition, occurring between 7.0 (micelles) and 6.2 (vesicles). Considering the pKa of 5.9 for RhaC10C10, the transition occurs when slightly more than 50% of RhaC10C10 molecules are deprotonated. The evolution between $\alpha$> 3 in the micellar region and $\alpha$< 2.4 in the vesicle region shows that micelles coexist with a small fraction (~ few % by number) of large-scale fractal structures, which eventually evolve together, and possibly merge into vesicles. Merging of micelles is also suggested by the increase in the core aspect ($X_{core}$= $L_{core}$/$R_{core}$) ratio when approaching the transition. However, the evolution from about 2 to 5 is still modest compared to what was found for the microbial glucolipid C18:1, also undergoing a micelle-to-vesicle transition.[42,57] In that system, micelles grow into giant wormlike cylinders, with aspect ratio of more than 100; disks merged into disks, which eventually close-up into vesicles. That mechanism was in line with the literature.[61] In the present RhaC10C10 system, the aspect ratio is more modest, probably suggesting that merging prevails over micellar growth.

According to cryo-TEM in Figure 5a,b, vesicles have a diameter of at least 100 nm, while their membrane thickness is in the order of few nm. It is then possible to describe vesicles as approximately flat objects at the length scale studied here by SAXS, whereas flat membranes should show a -2 ($\alpha$= 2) dependency of Log($I$)-Log($q$). Since the combination of fitted SAXS profiles and cryo-TEM does not leave any doubt about the vesicular nature of RhaC10C10 below pH 6, the slightly higher exponent ($\alpha$= 2.4) could be due to aggregation.

From a structural point of view, the size of the hydrophilic-hydrophobic regions is comparable between the *Em* and *Mm*. If, within the error, $R_{core}$ is in the same order of $L$ and $T_{shell}$ comparable with $T_{face}$, one still observes that $T_{shell}$ > $T_{face}$ and $L$ > $R_{core}$. This could be explained by the contribution of the carbonyl group in the $COO^-$ form, more hydrophilic and hydrated, to the shell region of the micelles, if compared to the COOH form, less hydrated and water-soluble and possibly contributing more to the hydrophobic core of the membrane in the vesicle phase.

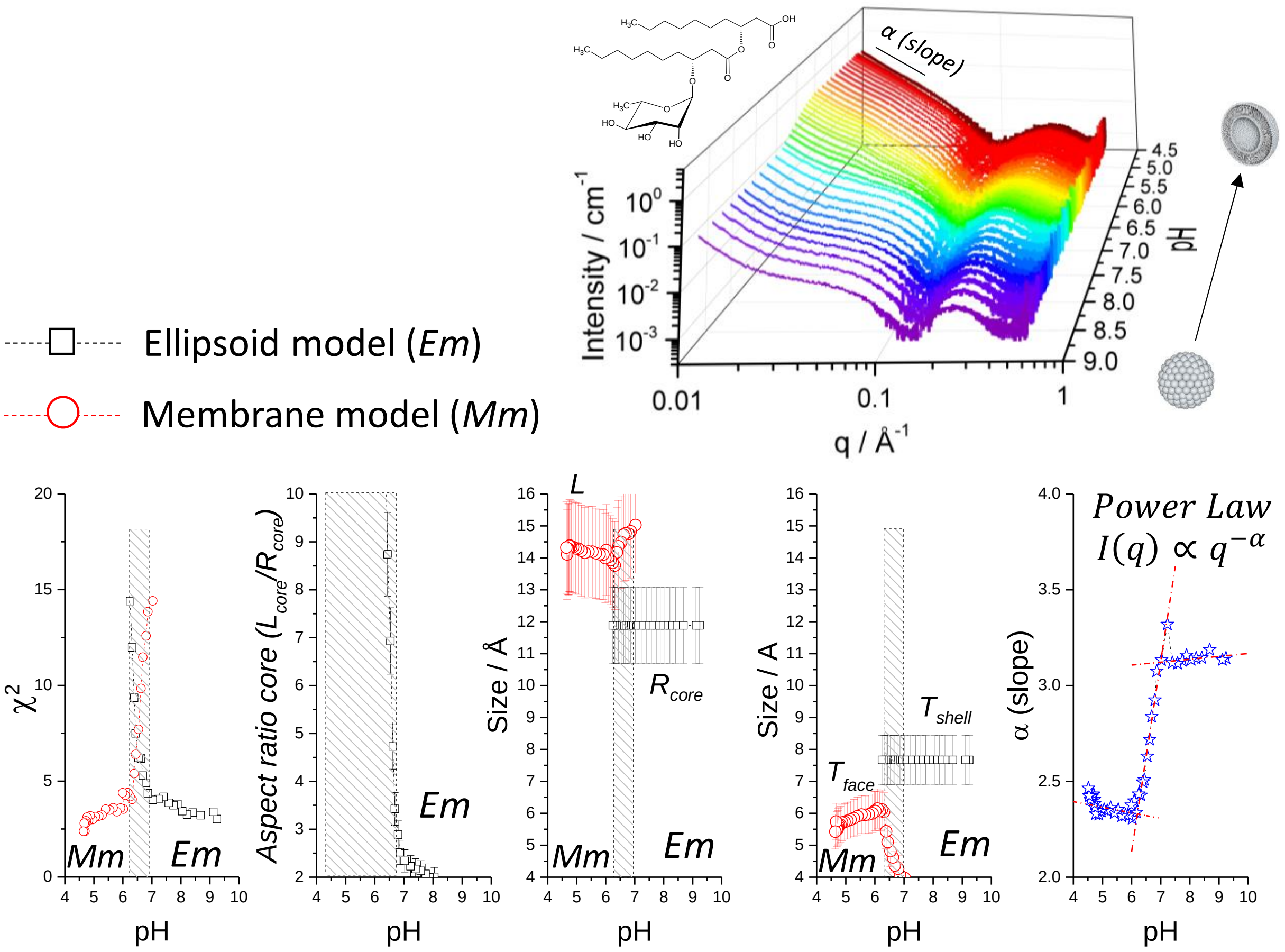

**Figure 7 – pH-resolved *in situ* SAXS experiment probing the micelle-to-vesicle transition of RhaC10C10 (5 mg/mL) and evolution of the related fitting parameters. The full list of the fit parameters is given in Table S 2 while details of the *Em* and *Mm* form factor models are given in Figure 2.**

*Quantitative analysis of SAXS data*

The self-assembly of RhaC10C10, RhaRhaC10C10 and their raw mixtures was studied in a both qualitative and quantitative way by a number of experimental and numerical approaches. Table 2 (from experiments) and Table 3 (from numerical simulations) summarize the main set of quantitative data extracted from the literature. Concerning the micellar phase, previous scattering data recorded for RhaRhaC10C10 and the commercial JBR515 (mixture of RhaC10C10 and RhaRhaC10C10) solutions above pH 7 provide a micellar radius contained between 15 Å and 20 Å. The data collected for RhaRhaC10C10 in this work nicely fall in this range. The total equatorial radius, ($R_{core}$ + $T_{shell}$), varies between 16.2 ± 1.6 Å at pH 7.5 and 19.1 ± 1.9 Å at pH 6.05. Similar values are found for RhaC10C10 at pH> 7.5 (19.6 ± 2.0 Å) and RhaC10 below pH 6 (17.8 ± 1.8 Å and 19.2 ± 1.9 Å). The good correlation also occurs for the specific core radius and shell thickness, reported to be 7.7 ± 0.1 Å and 12.0 ± 0.3 Å by Mortensen et al.[21] Concerning RhaC10, it certainly becomes more soluble and too hydrated to form a micelle with a well-defined core-shell structure at pH> 6. Finally, when available, the hydrodynamic radius ($R_h$) and the radius of gyration ($R_g$) are also in this same range, and generally not beyond 25 Å (Table 2).

The interesting aspect of comparing the micellar radius of RhaRhaC10C10, RhaC10C10 and RhaC10 is the fact that when these molecules assemble into micelles, they have the same total equaorial radius and with hydrophilic-hydrophobic regions of comparable size, despite the fact that the number of rhamnose units and C10 chains varies. In analogy to monosaccharides, the length of a single rhamnose unit should not exceed 10 Å, while a C10 chain has a length in the order of 12 Å, according to the Tanford formula, and considering only 8 $CH_2$ groups (1.54 + $n$1.265 Å, with $n$ being the number of methylene groups in the chain).[62] If added together, a rhamnose unit and a C10 tail should then have a total length of maximum 25 Å, considering the presence of the carbonyl. The experimentally-measured equatorial radius is rather in the order of 20 Å, independently of the number of rhamnose and C10 units, thus indicating that the micellar diameter is most likely constituted by two adjacent RLs, as classically found for classical head-tail surfactants. This is in contrast with a number of other bioamphiphiles, like sophorolipids or glucolipids,[42] for which the micellar diameter is equivalent to the length of the glycolipid itself, as expected for bolaamphiphiles.[63] Furthermore, RLs seem to form micelles with a well-defined core-shell structure, differently than acidic sophorolipids, which had shown a more complex "coffee bean-like" morphology, characterized by narrow central hydrophobic core and broader hydrophilic regions of less-defined composition.[43]

The number of rhamnose units and C10 tails affect neither the micellar shape nor its size nor its structure. This fact supposes that the rhamnose groups are all adjacent in the hydrophilic shell, independently of the mono or dirhamnose nature of the RLs and that each rhamnose group occupies a constant surface area. The consequence of this assumption is that, in order to map the micellar surface shell with rhamnose, the aggregation number of RhaC10 doubles that of RhaRhaC10C10, as shown in Table 4. This is is discussed in more details below.

**Table 2 – Literature survey of selected *experimental* structural parameters associated to RLs micellar and vesicular solutions. *C*= concentration, *S/RL*= surface area per RL molecule, $N_{agg}$= aggregation number, $R_g$= radius of gyration, $R_h$= hydrodynamic radius, *Th*= thickness of the vesicle membrane, *PP*= packing parameter.**

| C / mM | pH | *S/RL* / Å² | $N_{agg}$ | Phase | Radius / Å | | | *Th* / Å | *PP* | Ref |
|---|---|---|---|---|---|---|---|---|---|---|
| | | | | | From fit | $R_g$ | $R_h$ | | | |
| RhaC10C10 | | | | | | | | | | |
| | 9 | 52-77 | 47 | Micelles | | | | | 0.67 | 25 |
| | 9 | 52-77 | | Vesicles | | | | | 0.67 | 25 |
| 0-35 | 8 | 40 | | Micelles | | | 10-20 | | | 59 |
| | 8 | 117 | | | | | | | | 64 |
| | 6.8/7/8 | 86/109/98 | 26/31 (20 mM, pH 8) | Micelle (pH8) with minority vesicles (> 100 mM) | | | 24 (pH8) | | | 27 |
| <10 | 6-6.8 | 68 | | Micelles | | | | | 0.62 | 32 |
| >10 | 6-6.8 | | | Vesicles | | | | | | 32 |
| | 4 | 21 | | | | | | | | 64 |
| RhaRhaC10C10 | | | | | | | | | | |
| 20-100 | 9 | 77-80 | 26-86 | Micelles (ellipsoid) | 15-15 | | | | 0.5 | 25 |
| 33 | 7.4 | | | Micelles | 19.1 | | | | | 26 |
| 2 | 7.4 | | | Mix ves/micelle | | | | | | 65 |
| <100 | 6-6.8 | 56 | | Micelles | | | | | 0.73 | 32 |
| >100 | 6-6.8 | | | Vesicles | | | | | | 32 |
| 33 | 4.5 | | | Lamellae | | | | 27.9 | | 26 |
| RL mixture | | | | | | | | | | |
| [2 wt%] | 13.2 | | | Micelles | | 17.1 | | | | 20 |
| 7.6-12.7 | 7 buff | | 11.2 | Micelles (ellipsoid) | 19.7 | | | | | 21 |
| | | | | Vesicles | | | | 27 interdigi | | 22 |
| [> 80-200 mg/L ] | 7.4 buff | | | Vesicles | | | | | | 23 |
| [< 80 mg/L ] | 7.4 buff | | | Micelles | | | | | | 23 |
| 26 | 4 → 7 | | | Vesicle→ micelles | | | | | | 24 |
| [2 wt%] | | | | Vesicle | | | | 14 | | 20 |
| | variable | | | Vesicle-micelles | | | | | | 16,18 |

**Table 3 - Literature survey of selected *simulated* structural parameters associated to RLs micellar and vesicular solutions. *C*= concentration, *S/RL*= surface area per RL molecule, $N_{agg}$= aggregation number, $R_g$= radius of gyration, *Th*= thickness of the vesicle membrane, *PP*= packing parameter. *= anionic RLs.**

| C / mM | *S/RL* / Å$^2$ | Phase | Radius / Å | $N_{agg}$ | $R_g$ / Å | *Th* / Å | *PP* | Ref |
|---|---|---|---|---|---|---|---|---|
| RhaC10C10 | | | | | | | | |
| 118.7 mM (6 wt%) | 18.6 (calc) | Micelles | | 19 | 53 | | | [66] |
| 17-70 | | Micelles (sphere-to-wormlike) | | | 38 | | 0.61 | [29] |
| 70-140 | 111 (calc) | Vesicles | | 307 | 52 | | 0.61 | [29] |
| | 80 Å$^2$ | | | 40 | | | | [67] |
| 8.3 – 60 | 92 | Micelles | 12-25 | 5-44 | | | | [28] |
| 80 – 810 | | Vesicles | | | | 16 | | [28] |
| | | Pre-micellar/Micellar* | 10-40 | 25 most often (7-95 general) | | | | [27] |
| RhaRhaC10C10 | | | | | | | | |
| 118.7 mM (6 wt%) | 22.4 | Micelles | | 14 | 50 | | | [66] |
| 17-140 | | Micelles (sphere-to-wormlike) | | | | | 0.48 | [29] |
| [64/3500 H2O] | 69 | Membrane | | | | 31 | | [26] |
| | 90 | Spheres-cylinders | | 22 | | | | [67] |
| [64/3500 H2O] | 145 | Micelles* | | | | | | [26] |

Table 4, with Table S 3 and Table S 4 containing the upper and lower range limits respectively, report additional structural parameters calculated from geometrical assumptions. The micelle is assumed as being a prolate ellipsoid of revolution with the three semi-axes being *a* ($R_{core} + T_{shell}$), *b* ($R_{core} + T_{shell}$) and *c* ($R_{core}\, X_{core} + T_{shell}$). The values of the semi-axes are also defined in Table 4 and calculated from $R_{core}$, $T_{shell}$ and $X_{core}$ (Table 1). On this basis, one estimates the volume of the core, $V_{core}$ (Eq. 4) and the volume of the shell, $V_{shell}= V_{ellips}-V_{core}$, with $V_{ellips}$ given in Eq. 5. One can also measure the surface of the core, $S_{core}$ (Eq. 6), at the core-shell interface. Since the micellar shape is ellipsoidal, the surface is calculated with the Knud Thomsen's formula (Eq. 6). Given the above, the surface area per RL molecule, *Area/RL* (Eq. 7), the aggregation number, $N_{agg}$ (Eq. 8), and the number of water molecules per rhamnose group, $n_{H2O/Rha}$ (Eq. 9), and are calculated as follows:

$V_{core} = \frac{4}{3}\pi\ (R_{core})^3 X_{core}$ Eq. 4

$V_{ellips} = \frac{4}{3}\pi\ abc$ Eq. 5

$S_{core} = 4\pi \left(\frac{(RcoreRcore)^p + (Rcore(RcoreXcore))^p + (RcoreXcore)^p}{3}\right)^{1/p}$, $p \approx 1.6075$ Eq. 6

$Area/RL = S_{core}/N_{agg}$ Eq. 7

$N_{agg} = V_{core}/(xV_{C9})$ Eq. 8

$n_{H2O/Rha} = n_{H2O}/(yN_{agg})$ Eq. 9

In the calculation of $N_{agg}$, the hypothesis is made that the C9 backbone ($-(CH_2)_8CH_3$) of the C10 hydroxydecanoic acid is part of the hydrophobic core and the carboxylic acid is part of the hydrophilic shell. This approach is quite classical in estimating the aggregation number of surfactant micelles. In this context, $V_{C9}$ is the volume of the C9 backbone, calculated with the Tanford formula, V= 27.4 + 26.9*nC*, with *nC* being the number of methylene groups in the acyl chain (here, *nC*= 8). The sense of *x* is to account, in a single RL molecule, for two C10 groups for RhaC10C10 and RhaC10C10 (*x*= 2) and one single C10 for RhaC10 (*x*= 1). In the calculation of $n_{H2O/Rha}$, one employs $n_{H2O}$, the total number of water molecules in the hydrophilic shell, with *y* being the number of rhamnolipid groups in RhaRhaC10C10 (*y*= 2) and RhaC10C10, RhaC10 (*y*= 1). $n_{H2O}$ is the number of water molecules required to fill the hydrophilic region of the micelle, which is assumed to contain both rhamnose and COOH, so to match the total volume of the hydrophilic shell ($V_{shell}$), calculated from the SAXS analysis. $n_{H2O}$ is then estimated as follows (Eq. 10):

$n_{H2O} = [V_{shell} - (yV_{Rha} + xV_{COOH})]/V_{H2O}$ Eq.10

with $V_{Rha}$= volume of a single rhamnose group and $V_{COOH}$= volume of a carboxylic acid group, with *x* and *y* defined above. To estimate $V_{Rha}$ and $V_{COOH}$, we have employed the molar volume of rhamnose (113.89 $cm^3$/mol, that is 189 $Å^3$ per rhamnose, Table S 5) and, as first approximation, the molar volume of formic acid, FA (37.8 $cm^3$/mol, that is 63 $Å^3$ per FA, Table S 5).

**Table 4 – Derived structural parameters (defined in the main text) calculated for RLs micelles using the set of equations (Eq. 4 - Eq. 10 and Eq. 14 – Eq. 16 for *PP*) and molecular properties given in Table S 5. *a*, *b* and *c* represent the semi-axes of the ellipsoid of revolution model of Figure 2. The (6) and (8) subscripts correspond to a calculation for which 6 and 8 $CH_2$ groups in the C10C10 chain, respectively, are considered for the hydrophobic core of the micelles. Higher and lower limits are given in Table S 3 and Table S 4.**

| | **RhaRhaC10C10** | | **RhaC10C10** | **RhaC10** | | |
|---|---|---|---|---|---|---|
| pH | 6.05 | 7.5 | >7.5 | 3.97 | 5.11 | 6.02 |
| $a$ ($R_{core}$ + $T_{shell}$) / Å | 19.1 | 16.2 | 19.6 | 19.2 | 17.8 | 9.5 |
| $b$ ($R_{core}$ + $T_{shell}$) / Å | 19.1 | 16.2 | 19.6 | 19.2 | 17.8 | 9.5 |
| $c$ ($R_{core}$ $X_{core}$ + $T_{shell}$) / Å | 36.5 | 28.2 | 37.5 | 37.3 | 30.3 | 58.9 |
| $V_{core}$ / Å$^3$ | 7421 | 5359 | 17638 | 8255 | 5985 | 8648 |
| $V_{shell}$ / Å$^3$ | 48373 | 25626 | 42595 | 49251 | 34142 | 13611 |
| $S_{core}$ / Å$^2$ | 2188 | 1671 | 3697 | 2349 | 1798 | 3334 |
| *Area/RL* / Å$^2$ | 143 | 151 | 102 | 69 | 73 | 94 |
| $n_{H2O/Rha(8)}$ | 2.9 | 2.7 | 0.9 | 1.2 | 1.5 | 0.1 |
| $n_{H2O/Rha(6)}$ | 2.0 | 1.9 | 0.6 | 0.8 | 1.1 | 0.0 |
| $N_{agg(8)}$ | 15 | 11 | 36 | 34 | 25 | 36 |
| $N_{agg(6)}$ | 20 | 14 | 47 | 44 | 32 | 46 |
| $PP_{(8)}$ | 0.41 | 0.40 | 0.40 | 0.41 | 0.40 | 0.43 |
| $PP_{(6)}$ | 0.32 | 0.31 | 0.31 | 0.32 | 0.31 | 0.33 |

It is found that the aggregation number, $N_{agg(8)}$ (Table 4) is contained between 10 and 15 for RhaRhaC10C10, 36 for RhaC10C10 and between 25 and 36 for RhaC10. Compared to literature, and within the higher and lower limits estimated from the 10% error on the parameters of the fit (Table S 3 and Table S 4), these values are in the same order of magnitude of those reported for both RhaRhaC10C10 (26-86) and RhaC10C10 (26-31, 47) (Table 2). Values of aggregation number as low as 11 were actually reported by Mortensen et al. using a similar SAXS modelling approach.[21] Values in the order of 11 were also reported for the commercial JBR215 compound (Table 2). Even if there is agreement between the aggregation numbers found here and literature, values close to 10 still seem relatively low compared to the majority of data reported for RLs (Table 2). One possible source of error in the estimation of $N_{agg}$ could be attributed to the hypothesis that all methylene groups (16 for RhaRhaC10C10 and RhaC10C10 and 8 for RhaC10) are located in the hydrophobic core. Considering that former studies on surfactant micelles have shown that methylene groups close to the hydrophilic headgroup are hydrated and could be part of the hydrophilic shell,[45–47] one could then estimate new values of $N_{agg}$ with the hypothesis that the core only contains 12 $CH_2$ for C10C10 RLs and 6 for C10 RLs. The new aggregation numbers are expressed as $N_{agg(6)}$ in Table 4, where the subscript *(6)* indicates the number of $CH_2$ groups of the C10 in the core. $N_{agg(6)}$ are about 30% higher than $N_{agg(8)}$ and are still in full agreement with the literature values (Table 2), thus suggesting that hydration of part of the C10 backbone is not to be excluded.

Another interesting parameter, which could also serve as a control, is the hydration number of each rhamnose group, expressed in Table 4 as $n_{H2O/Rha(8)}$ and $n_{H2O/Rha(6)}$ for the 8-$CH_2$ and 6-$CH_2$ hypotheses, respectively. In the 8-$CH_2$ hypothesis, the number of water molecules per rhamnose group is about 3 for RhaRhaC10C10 and less than 1.5 for the other RLs. RhaC10 above its pKa value represents the only exception, because its hydration number cannot be estimated, most likely due to its poorly-defined micellar nature at pH> 6. In the 6-$CH_2$ hypothesis, $n_{H2O/Rha(6)}$ is about 30% smaller than $n_{H2O/Rha(8)}$ (the upper and lower limits are given in Table S 3 and Table S 4). Although it is not an easy task to determine the hydration number of carbohydrates, the values obtained for the 6-$CH_2$ hypothesis and falling in the range of 1, or less, for most systems in Table 4, seem exceedingly low. Rare are those studies, which measured the hydration number of RLs. Euston et al. have estimated it using molecular dynamics and proposed numbers in the range of tens of water molecules, most likely associated to the carbonyl group than to rhamnose.[68] The values reported by Euston et al. seem, on the other hand, exceedingly high compared to what has been reported for other systems and suggest that they probably consider more than one hydration layer, thus falling out of the hydration shell probed by SAXS. Their data recall more the hydration number estimated by Winther et al.,[69] who used a geometric hydration number, identified as the number of water molecules required to cover the solute with one layer (47 water molecules per trehalose). The actual number of water molecules per sugar unit is most likely much smaller. Simulation data performed on RhaC10C10 provide 3 water molecules per RL in the case of micelles and 2.2 in the case of vesicles.[28] In the case of maltoside surfactants, the hydration number measured by SAXS was more seemingly estimated to 8.[70] In the case of single sugars, authors have estimated the hydration number of trehalose to be between 4 and 18,[71] while in the case of glucose, hydration number can vary between 0 and 15,[72] depending on conformation. Considering the upper and lower limits estimated by the 10% error of the SAXS fitting process, it seems to us that the hydration numbers estimated by the 8-$CH_2$ hypothesis are most likely more realistic.

Finally, the surface area associated to a RL molecule is calculated at the core-shell interface ($S_{core}/N_{agg}$). One finds values are in the range 130-170 $Å^2$ for RhaRhaC10C10, around 100 $Å^2$ for RhaC10C10 and between 60 and 100 $Å^2$ for RhaC10 (Table 4). The experimental literature values of the surface area per RL molecule are quite sparse (Table 2). For RhaC10C10, values ranging between 40 Å$^2$ [59] and 117 Å$^2$ [64] and up to 135 Å$^2$ [73] were reported above the pKa, while these values vary between 21 Å$^2$,[64] 68 Å$^2$ [32] and 77-80 Å$^2$ [25] below pH 7. For RhaRhaC10C10, one finds values ranging around 80 Å$^2$,[25] 56 Å$^2$ [32] and up to 131 Å$^2$.[73] The values estimated from numerical modelling are reported in Table 3 and are just as sparse, ranging from 18.6 Å$^2$ to 111 Å$^2$ for RhaC10 and from 22.4 Å$^2$ to 145 Å$^2$ for RhaRhaC10C10. In the broader literature of glycolipids, one also finds an area of 50 Å$^2$ at the core/sugar headgroup interface for dodecyl maltoside surfactants[74] and values ranging between 64 to 79 Å$^2$ for sophorolipid molecules at the core/shell interface.[43]

The literature survey shows that the range of values of the surface area per molecule measured for RLs is wide, and it seems to depend a lot on the medium (pH, buffer, ionic strength).[27] Comparison between the data estimated in this work and literature indicates that the values found here fall in the upper range, with an excess in the case of RhaRhaC10C10. The possible discrepancy could be explained by the method of evaluation. Here, the surface area is derived from geometrical assumptions after fitting the SAXS profiles. The advantage of this approach is to estimate the surface area in the micelle itself, while the drawback is the high error (±10%) correlated to the fitting process and related propagation in the geometrical calculations. On the contrary, (experimental) surface area in the literature is generally obtained by surface tension experiments combined to the Gibbs adsorption isotherm equation. This method reduces the sources of error, but it gives access to a surface area measured at the planar air-water interface and reflecting a tighter packing of the molecules. Despite the difference with the surface tension method adopted in the literature, the present approach provides values of the packing parameter, which are compatible with the shape found experimentally, as outlined in more detail below.

*Structure of the membrane in RhaC10C10 vesicles*

RhaC10C10 at pH< 6.5 undergoes a micelle-to-vesicle transition (Figure 7), with evolution of structural parameters given in Figure 7 and Table 1. According to the data, the total thickness of the vesicle membrane, $T_{total}$, composed of RhaC10C10 is, $T_{total} = 2T_{face} + L$. Depending on pH (Figure 7), 24.9 Å < $T_{total}$ < 26.3 Å. As previously discussed, the expected length of a RhaC10C10 molecule, according to the Tanford formula applied to the $CH_3(CH_2)_8$ chain[62] and typical length of monosaccharides, should not exceed 25 Å. By comparing $T_{total}$ with the expected length of RhaC10C10, it seems reasonable to state that $T_{total}$ is in the order of the full RL length. The vesicle membrane is then composed of an interdigitated single layer of the RL, rather than a bilayer, as classically found for phospholipids and in good agreement with several work on RLs (Table 2 and Table 3).[22,26]

If one employs the same hypothesis as for micellar aggregates, that is the assumption that the hydrophobic region of the membrane is in a liquid crystalline state, one can evaluate the surface area per RL at the core-shell interface in membranes, $\frac{Area}{RL}_{memb}$ (Eq. 11) as follows:

$$\frac{Area}{RL}_{memb} = \frac{xV_{C9}}{L} \qquad \text{Eq.11}$$

derived from Eq. 12 and Eq. 13,

$$\frac{Area}{RL}_{memb} = \frac{S_{memb}}{N_{agg-memb}} = \frac{S_{memb}xV_{C9}}{V_{memb-core}} \qquad \text{Eq.12}$$

$$N_{agg-memb} = \frac{V_{memb-core}}{xV_{C9}} \qquad \text{Eq.13}$$

with $N_{agg\text{-}memb}$ (Eq. 13) being the aggregation number of RLs in the membrane, $V_{memb\text{-}core}$ being the volume of the core region and considered here as the volume of a parallelepiped of arbitrary side dimensions, *y*,*z*, and thickness, *t* and with $S_{memb}$ being its surface, given by the product of *yz*. In the present system, given the interdigitation of RL molecules, $t \equiv L$, with *L* given in Table 1, and *x*= 4, as it accounts for two RhaC10C10 molecules in the hydrophobic layer (interdigitation), each carrying two C10 groups of volume $V_{C9}$ (243 Å$^3$ according to Tanford in the 8-$CH_2$ hypothesis). Altogether, the value of $\frac{Area}{RL}_{memb}$ depends on *L*, which only varies slightly with pH (Table 1), thus giving a surface area, $\frac{Area}{RL}_{memb}$ = 70 ± 7 Å$^2$, the error being associated to the uncertainty of the fitting process (± 10%). This value is in very good agreement with the range (65 Å$^2$ to 135 Å$^2$)[73] determined for RhaC10C10 (Table 2) by surface tension at the air-water interface. Good agreement also occurs with the value of 69 Å$^2$ obtained from simulation of RhaRhaC10C10 vesicles (Table 3).[26]

*Relationship between RL structure and properties*

On the basis of the experimental structural data collected from SAXS for RhaC10, RhaC10C10 and RhaRhaC10C10, one must acknowledge very good agreement between the values of $PP$ and the morphology of the self-assembled structures. A simplistic overview of the structure-property relationship for the three RLs studied in this work is given in Figure 8, with the corresponding quantitative structural data given in Table 1, Table 4 and Figure 7. To better correlate the molecular structure of RLs and the morphologies of their self-assembled structures, we employ the notion of packing parameter, $PP$.

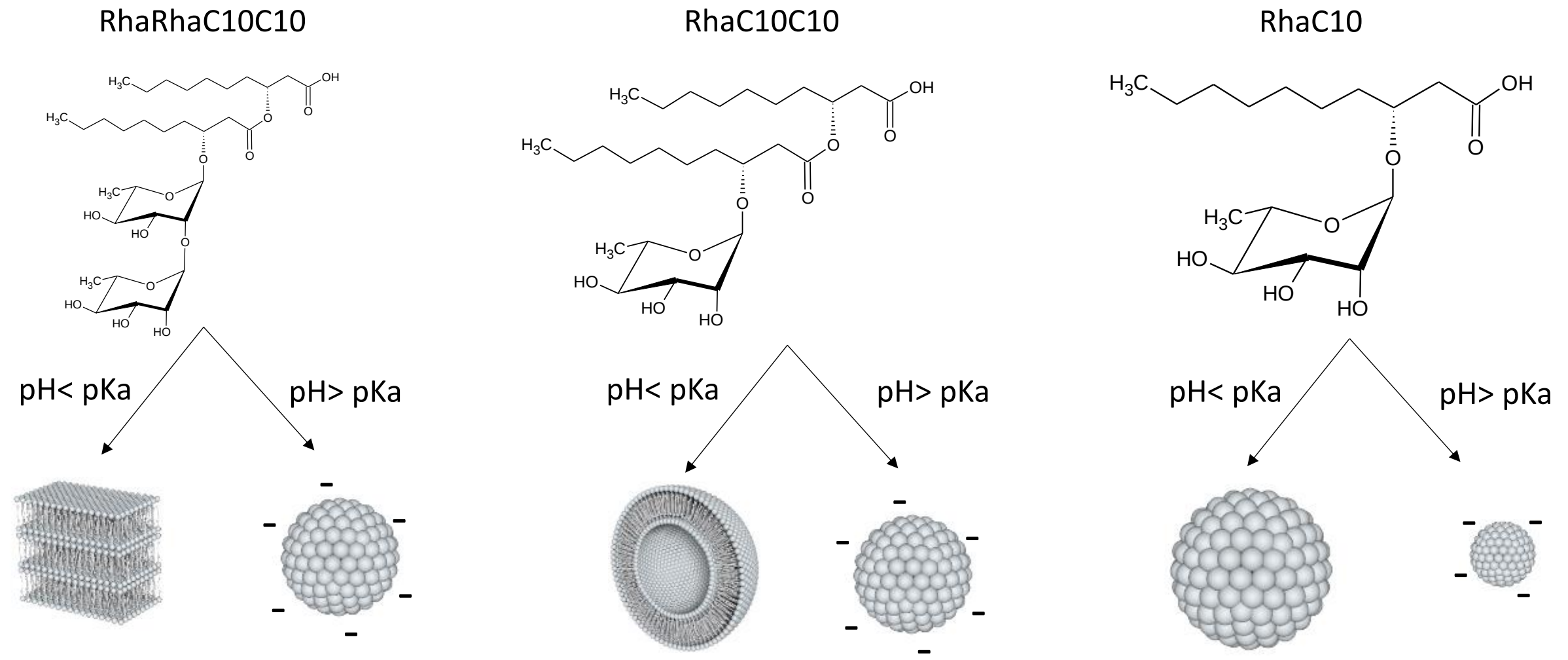

**Figure 8 – Typical pH-dependent phase behavior in water at concentration up to 25 mg/mL of rhamnolipids RhaRhaC10C10, RhaC10C10 and RhaC10. Drawings of micelles, lamellae and vesicle are not scaled. Drawing only are meant to represent a relative change in morphology (RhaRhaC10C10 and RhaC10C10) or a change in size (RhaC10).**

The $PP$ theory was developed in the 1970s to explain and predict the self-assembly of amphiphiles.[30,75] $PP$ is commonly expressed as the ratio between the volume ($V_{amph}$) of the amphiphile's hydrophobic chain and the product of its length ($L_{amph}$) with the equilibrium surface area ($A_e$) at the core-shell interface (Eq. 14).

$$PP = \frac{V_{amph}}{L_{amph}A_e} \qquad \text{Eq.14}$$

In the present systems, micellar or vesicular, if one assumes that $V_{amph} \equiv xV_{C9}$ ($x$ being the number of C10 chains in a RL molecule, with $V_{C9}$ being 243 Å$^3$), $L_{amph} \equiv R_{core}$ for micelles or $L_{amph} \equiv L$ for vesicles and $A_e \equiv Area/RL$ for micelles (or $A_e \equiv \frac{Area}{RL}_{memb}$ for vesicles), it is possible to determine the experimental $PP$ for RL micelles (Eq. 15) and vesicles (Eq. 16):

$$PP = \frac{xV_{C9}}{R_{core}\,Area/RL}\ \text{(micelles)} \qquad \text{Eq.15}$$

$$PP = \frac{xV_{C9}}{L\frac{Area}{RL}_{memb}}\ \text{(vesicles).} \qquad \text{Eq.16}$$

In the case of micellar systems ($x$= 1 for RhaC10 and $x$= 2 for RhaC10C10 and RhaRhaC10C10; $R_{core}$ and *Area/RL* given in Table 1 and Table 4), the values of experimental *PP* calculated for the 8-$CH_2$ and 6-$CH_2$ hypotheses are 0.41 ± 0.01 and 0.32 ± 0.01, respectively. The error corresponds to the standard deviation calculated over all *PP* data given in Table 4, while the error associated to the uncertainty (± 10%) of the fitting process does not influence at all these values (Table S 3 and Table S 4). For vesicles (RhaC10C10 at pH< 6.5), with $L$ given in Table 1 and $\frac{Area}{RL}_{memb}$ = 70 ± 7 $Å^2$, one finds 0.44 < *PP* < 0.55, considering the uncertainty. For the lamellar phase observed for RhaRhaC10C10 at acidic pH, *PP*= 1 by definition.

These experimental *PP* values (0.32-0.41 for micellar RLs and 0.44-0.55 for vesicular RLs) are in very good agreement with the experientally-found morphologies. For *PP*< 1/3, one expects spherical micelles, for 1/3 < *PP* < 1/2, one expects ellipsoidal/cylindrical micelles, and for 1/2 < *PP* < 1, one expects vesicles. Interestingly, the match between the experimental value of *PP* and the observed morphology is better than what others have found in the literature using the value of area per RL, $A_e$, extracted from surface tension experiments. As shown in Table 2, values of *PP* above 0.5 are generally proposed for micellar systems. In fact, the discrepancy between the calculated *PP* and the actual observed morphology was specifically discussed. Chen *et al.*[25] found good agreement between experiments and theory (*PP*> 0.5) for RhaC10C10 (membranes) but acknowledged that theory (*PP*> 0.5) failed to explain the presence of micelles for RhaRhaC10C10 (*PP*< 0.3/0.5). They attributed the origin of the problem to the methodology (surface tension) employed to estimate $A_e$ and $L_{amph}$.

To better understand the impact of the RL structure on its self-assembly, Table 5 compares the *PP* values calculated from theoretical and experimental approaches. For the theoretical *PP*, $L_{amph}$= 14 Å (Tanford) and $V_{amph}$= 243 $Å^3$ for RhaC10 and 485 $Å^3$ for RhaC10C10 and RhaRhaC10C10. Two set of values were used for $A_e$. The first set (* column) contains the values measured by Chen *et al.* using surface tension (more detail in the caption of Table 5).[25] The second set (** column) employs the value of 70 $Å^2$ (determined by SAXS on RhaC10C10 at pH< 6.5 in this work) for RhaC10C10 and RhaC10 and the value of 140 Å$^2$ for RhaRhaC10C10, assuming a double surface area due to the di-rhamnose group.

**Table 5 – Calculation of packing parameter (*PP*) of rhamnolipids employed in this work using Eq. 14 – Eq. 16. The theoretical *PP* is calculated using molecular data in Table S 5 and using the Tanford formula. The experimental *PP* is calculated using structural data given in Table 1, Table 4 and Table S 5.**

| Molecule/conditions | Theoretical PP | | Experimental PP | Experimental Structure | Agreement between theory and experiment |
|---|---|---|---|---|---|
| RhaRhaC10C10/pH 5.0 | 0.45* | 0.25** | - | Lamellae (*PP*= 1) | No |
| RhaRhaC10C10/pH 6.05 | 0.45* | 0.25** | 0.32-0.41 | Micelle | Good |
| RhaRhaC10C10/pH 7.5 | 0.43* | 0.25** | 0.32-0.41 | Micelle | Good |
| RhaC10C10/pH > 7.5 | 0.45* | 0.50** | 0.32-0.41 | Micelle | Poor |
| RhaC10C10/pH < 6.5 | 0.53* | 0.50** | 0.44-0.55 | Vesicle | Good |
| RhaC10/pH 3.97 | 0.26* | 0.25** | 0.32-0.41 | Micelle | Good |
| RhaC10/pH 5.11 | 0.26* | 0.25** | 0.32-0.41 | Micelle | Good |
| RhaC10/pH 6.02 | 0.26* | 0.25** | 0.32-0.41 | Micelle | Good |

*RhaC10*. Independently of the method of calculating the surface area and of pH, the RhaC10 molecule is always expected to form micelles in water. This is indeed experimentally verified at all pH tested, representative of the protonated and deprotonated state of RhaC10. The *PP* approach is then well-suited to explain and predict the self-assembly of this compound. The *PP* calculated from experimentally-measured structural parameters is certainly higher than the theoretical *PP*, but still in the range of value expected for micellar aggregates. This being said, one should notice the evolution of the surface area per molecule with increasing pH: from 69 Å$^2$ (protonated) to 94 Å$^2$ (deprotonated) (Table 4). Interestingly, the value of 70 Å$^2$ is also measured for the protonated RhaC10C10 in the vesicle phase. This suggests that the excess area of 25 Å$^2$ at higher pH should be attributed to the contribution of the carboxylate group.

*RhaC10C10*. This compound has a double fatty acid chain, thus doubling the hydrophobic volume, but constant tail length and surface area (70 Å$^2$, estimated for vesicles) in the protonated regime (low pH) compared to RhaC10. The effect is the increase in the *PP* to values in the order of 0.5, expected for vesicles, which are indeed experimentally found at pH< 6.5. On the contrary, deprotonation of the carboxylic acid at pH> 7.5 is responsible for increasing the surface area per RL from 70 Å$^2$ to 102 Å$^2$. The excess of 32 Å$^2$, comparable with what it was found in the case of RhaC10 and due to electrostatic repulsion, is enough to lower the experimental *PP* from 0.44-0.55 to 0.31-0.41, thus explaining the vesicle to micelle transition observed when increasing pH. The surface area excess due to the contribution of the carboxylate group also explains the poor agreement between the experimental and theoretical *PP*, for which the value of 77 $Å^2$ [25] was employed. Of course, values of surface area as high as 117 $Å^2$ (Table 2) or 135 $Å^2$ [73] reported before would reduce the actual *PP*, thus improving the prediction. However, the problem is precisely the difficulty to measure a reliable value of surface area of RhaC10C10, as shown by the broad range existing in the literature (Table 2 and Ref. [27]).

*RhaRhaC10C10*. This molecule has an additional rhamnose group, which has the tendency to increase the surface area, thus reducing the theoretical *PP*. In a crude approach, which consists in considering a surface area of 140 $Å^2$, the double of a single rhamnose taken at acidic pH, thus excluding electrostatic effects, the *PP* falls below 0.3. If the value of 77 $Å^2$, measured by Chen et al.[25] at pH 7 using surface tension, is used instead, *PP* is rather in the order of 0.45. In both cases, *PP* is in agreement with the experiment (micellar morphology). However, the PP approach fails at low pH, in the protonated state of RhaRhaC10C10, which precipitates in a lamellar phase (Figure 3 and Ref. [26]), for which *PP* is 1 by definition. Unfortunately, unless specific pressure-distance experiments are performed, it is not possible to measure the structural (thickness of the membrane, core length) parameters of the lamellar structure, due to the uncertainty in the size of the interstitial water layer.

To explain the assembly of RhaRhaC10C10 in a flat geometry using the *PP*, one should consider the possibility of an effective shorter C10 chain (~ 8 Å) and a small surface area (~ 60 $Å^2$), while keeping the volume of the double C10 chain constant. To account for such unpredictable structural parameters, one can only suppose an important conformational rearrangement of the di-rhamnose group, which would consist in a tight packing of the di-rhamnose group along their longitudinal axis, so to minimize the surface area per molecule. Possible tilting, or bending, of the C10C10 moiety is not excluded, to account for an effective shorter chain.

Conformational changes in glycolipids are not unusual. They were reported for alkylaldonamides,[76–78] with a critical impact on the self-assembly, but also more recently by us on bolaform di-sophorolipids[79] and glucolipids.[80] In the case of RLs, the amount of experimental data is limited, but a body of work, generated by numerical modelling,[26–28,67,68,81] agrees on the importance and impact of rhamnose headgroup conformation and C10C10 chain tilting on the self-assembly properties. Simulation data, which were generated for systems both at air-water interface[67,68,81] and in bulk water,[26–28] are sometimes contradicting, depending on the dilution regime,[67,68] and for this reason they must be considered with caution. Nevertheless, they all agree on the fact that intramolecular H-bonding between the rhamnose headgroup and the carbonyl function play a critical role in the conformation of both the rhamnose and chain in single-rhamnose lipids (RhaC10C10), while the second rhamnose group seems to be less involved in RhaRhaC10C10. Typical open, partially open and closed forms of RLs were reported,[28,67,68,81] justifying surface area per RL in the order of 60 Å$^2$ for RhaRhaC10C10 membranes.[26] Similarly, possible bending of the C10C10 chain was proposed to provide a length of less than 5 Å.[28]

The analysis above shows that, compared to other microbial glycolipids (see Table 9 in Ref. [17]), the theory of *PP* can be nicely applied to understand and predict the structure-property relationship in RLs. This is probably due to the somewhat standard head-tail configuration of this family of molecules compared to others, like sophorolipids or cellobioselipids (bolaamphiphiles), despie the presence of the free carboxylic acid. Nevertheless, RLs display the intrinsic, yet unpredictable, complaxity of many glycolipids, in relationship to the possible multiple conformations of the sugar headgroup, especially in the di-Rhamnose form. Considering the importance of RLs as biological amphiphile, further studies, correlating predictions and experiments, are still needed to achieve full control and understanding of this class of compounds.

**Conclusions**

In this work, three RLs with reasoned variation of their molecular structure were studied in water above their cmc as a function of pH by cryo-TEM and SAXS.

RhaC10 forms micelles (core-shell prolate ellipsoids) in a broad pH range, from acidic to basic. At basic pH, the micelles are small ($R_{core}$= 6.1 Å, $T_{shell}$= 3.4 Å) and most likely hydrated, with a poorly defined hydrophilic-hydrophobic contrast. At acidic pH, micellar boundaries are better defied and are larger in size ($R_{core}$= 8.3 Å, $T_{shell}$= 9.5 Å). The aggregation number varies between 25 and 36 and the area per RL increases from 69 Å$^2$ to 94 Å$^2$ when pH increases from 3.97 to 6.02. The theoretical PP for RhaC10 is 0.25, while the experimental PP is in the range 0.32-0.44. Both are compatible with the micellar structure observed experimentally. All in all, the pH does not seem to play a major role in the self-assembly of RhaC10, of which the self-assembly seems to be controlled by its shape.

RhaC10C10 forms micelles at basic pH and vesicles at acidic pH, the transition pH being around 6.5. At basic pH, micelles are best described as core-shell prolate ellipsoids with $R_{core}$= 11.9 Å and $T_{shell}$= 3.4 Å. The morphological transition towards vesicles is continuous with a possible elongation and merging. The membrane is best described with a core-shell structure, as well, with $T_{face}$ about 6 Å and core length of 14 Å. These parameters do not vary with pH. The aggregation number in the micellar phse is 36 and the area per RL decreases from 102 Å$^2$ (micelles) to 70 Å$^2$ (vesicles). The theoretical PP for RhaC10C10 is contained between 0.45 and 0.5, when calculated with surface area values collected from the literature. The experimental PP, calculated with surface area values from this work, is between 0.44 and 0.55 in the vesicle phase and 0.32 and 0.41 in the micelle phase. This shows that the agreement between the theoretical PP and the actual morphology is good for the vesicle phase but rather poor to predict the micelle phase. Employing surface area values measured directly on the micelle and vesicle phases improves the accuracy of the PP model. Finally, the discrepancy between the surface area per RL measured at acidic pH (COOH form) and basic pH (COO- form) for both RhaC10 and RhaRhaC10C10 shows that the contribution to the surface area attributed to the negatively-charged carboxylate group is between 25 and 35 Å$^2$. On the basis of the PP model, RhaC10C10 is expected to self-assemble into vesicles in its protonated form. Deprotonation introduces and extra contribution to the surface area of RL, which lowers the PP and it explains the vesicle-to-micelle transition.

Finally, RhaRhaC10C10 forms micelles in a broad pH range but it precipitates into a lamellar powder at pH below 5. Micelles are best described as core-shell prolate ellipsoids with $R_{core}$= 8.0 Å and $T_{shell}$ between 8 Å and 11 Å, aggregation number between 11 and 15 and area per RL between 143 Å2 and 151 Å$^2$. The theoretical PP for RhaRhaC10C10 using values of the surface area from the literature is contained between 0.25 and 0.45, while the experimental PP, calculated with surface area values from this work, is between 0.32 and 0.41. This shows a reasonably good agreement of both theoretical and experimental PP with the actual micellar phase observed at pH above 6, that is when RharhaC10C10 is partially deprotonated. On the other side, the agreement badly fails for the fully protonated form of the molecule, for which lamellar aggregates are observed (PP is unitary by definition).

The case of the acidic form of RhaRhaC10C10 shows the limits of the PP model in predicting and describing the self-assembly of RLs. In particular, to account for the smaller surface area necessary to explain the lamellar phase, one has to consider the possibility that the double rhamnose group undergoes important conformational changes between basic and acidic pH. It is not excluded that the C10C10 backbone folds to reduce the effective length. Conformational effects are not rare in glycolipids and are known to have a strong impact on the phase behaviour. Considering the fact that they are not taken into consideration in the PP model and they generally do not occur for classical surfactants, it illustrates the complexity in predicting the phase behaviour not only for RLs but for bioamphiphiles in general.

**Supporting Information**

Table S 1 presents the HPLC for RhaC10, RhaC10C10, RhaRhaC10C10 and C10C10 molecules, Table S 2 gives the full list of the fixed and variable parameters employed in the fitting of the SAXS profiles, Table S 3 and Table S 4 give the higher and lower limits, respectively, of the direct and derived structural parameters obtained by fitting the SAXS profiles of RL solutions, Table S 5 provides the structural molecular parameters employed to calculate molecular volumes.

Figure S 1 presents the surface tension experiments, Figure S 2 shows the contributions of the *Em*, *Mm* and *Power law* models to the fit of the entire SAXS profile, Figure S 3 shows the turbidity of C10C10 solutions at different pH values, Figure S 4 shows the fits of the SAXS profiles of RhaRhaC10C10 solutions at pH 7.5 and pH 6.05, Figure S 5 gives the fits of the SAXS profiles of the RhaC10 solution at pH 3.97, Figure S 6 shows a series of SAXS profiles of RhaC10C10 solutions extracted from the pH-resolved *in situ* experiment.

**Acknowledgements**

ESRF (proposal N. MX2311 and SC-5125) and Soleil (proposal N. 20201747) synchrotron facilities are kindly acknowledged for financial support during the beamtimes. NB and AP kindly acknowledge the French ANR, Project N° SELFAMPHI - 19-CE43-0012-01. CM and TT were partially funded by the Ministry of the Environment of the state of NRW (MULNV) within the framework of the "Sonderprogramm Umweltwirtschaft" in the project *RhamnoLizer*. CCB and TT have received funding from the Fachagentur Nachwachsende Rohstoffe (FNR) of the German Federal Ministry of Food and Agriculture (BMEL) within the project *KERoSyn*. TT and LMB acknowledge additional funding by the German Research Foundation (DFG) under Germany's Excellence Strategy within the Cluster of Excellence FSC 2186 "The Fuel Science Center".

# Supporting information

## Self-Assembly of Rhamnolipids Bioamphiphiles: Understanding Structure-Properties Relationship using Small-Angle X-Ray Scattering

Niki Baccile,[a,*] Alexandre Poirier,[a] Javier Perez,[b] Petra Pernot,[c] Daniel Hermida-Merino,[d,e] Patrick Le Griel,[a] Christian C. Blesken,[f] Conrad Müller,[f] Lars M. Blank,[f] Till Tiso[f]

[a] Sorbonne Université, Centre National de la Recherche Scientifique, Laboratoire de Chimie de la Matière Condensée de Paris, LCMCP, F-75005 Paris, France

[b] Synchrotron Soleil, L'Orme des Merisiers, Saint-Aubin, Gif-sur-Yvette, France

[c] ESRF – The European Synchrotron, CS40220, 38043 Grenoble, France

[d] Netherlands Organisation for Scientific Research (NWO), DUBBLE@ESRF BP CS40220, Grenoble, 38043, France

[e] Departamento de Física Aplicada, CINBIO, Universidade de Vigo, Campus Lagoas-Marcosende, Vigo, 36310, Spain

[f] iAMB – Institute ofApplied Microbiology, ABBt – Aachen Biology and Biotechnology, RWTH Aachen University, Aachen, German

* Corresponding author:

Dr. Niki Baccile

E-mail address: niki.baccile@sorbonne-universite.fr

Phone: +33 1 44 27 56 77

**Table S 1 – Retention times in HPLC known for RhaC10, RhaC10C10, RhaRhaC10C10 and C10C10 molecules. Typical chromatograms for each molecules and raw mix of samples are given below the table.**

| Congener | Retention time / min |
|---|---|
| RhaC10 | 2.5 |
| RhaRhaC10C10 | 5.1 |
| RhaC10C10 | 7.1 |
| C10C10 | 9.3 |

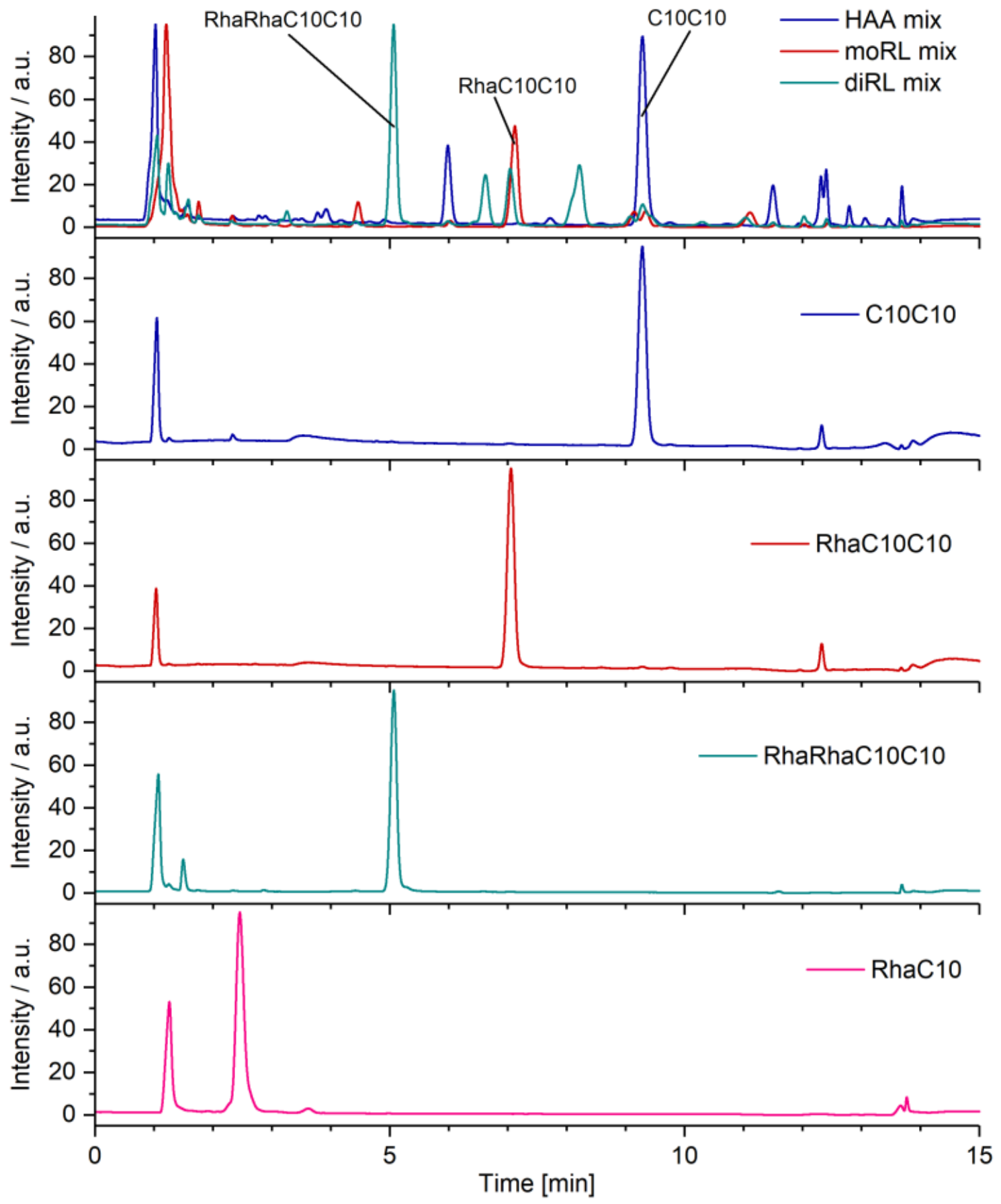

**Table S 2 – Full list of the fixed and variable parameters employed in the fitting of the SAXS profiles given in Figure 3 and Figure 7 using models (Figure 2) available in the SasView 3.1.2 software.**

## Sample : RhaRhaC10C10

Fixed parameters

| | **Core-Shell Ellipsoid (*Em*)** | | | | | **Power Law** | | | **Hayter MSA (structure factor)** | | |
|---|---|---|---|---|---|---|---|---|---|---|---|
| pH | *bkg* | *scale** | $\rho_{core}$ / Å$^{-2}$ | $\rho_{shell}$ / Å$^{-2}$ | $\rho_{solvent}$ / Å$^{-2}$ | *bkg* | *α* | Scale | T / °C | Dielectric | φ (scale) |
| 6.05 | $4.4.10^{-4}$ | 0.025 | $8.40 \cdot 10^{-6}$ | $9.9.0 \cdot 10^{-6}$ | $9.40 \cdot 10^{-6}$ | 0 | 4 | $2.10^{-4}$ | - | - | - |
| 7.5 | $4.4.10^{-4}$ | 0.025 | $8.40 \cdot 10^{-6}$ | $10.0 \cdot 10^{-6}$ | $9.40 \cdot 10^{-6}$ | 0 | 4.7 | $1.20^{-4}$ | 300 | 71.1 | 0.025 |

* the *scale* in the *Em* is set equal to the sample's volume fraction.

Variable parameters

| pH | Model | $R_{core}$ / Å | $T_{shell}$ / Å | $X_{core}$ | $X_{shell}$ | $R_{Hayter}$ / Å | *charge* |
|---|---|---|---|---|---|---|---|
| 6.05 | 1 | 8.3 | 10.8 | 3.1 | 1 | | |
| 6.05 | 1 | 8.0 | 11.6 | 3.9 | 0.5 | | |
| 6.05 | 1 | 7.8 | 12.2 | 4.8 | 0.1 | | |
| 7.5 | 2 | 8.0 | 8.2 | 2.5 | 1 | 19.3 | 6.4 |
| 7.5 | 2 | 7.4 | 9.6 | 5.2 | 0.1 | 22.1 | 5.6 |

## Sample : RhaC10C10

pH-resolved *in situ* experiments

| **Full model** | | | **Core-Shell Ellipsoid (*Em*)** | | | | | | | | | **Power Law** | | |
|---|---|---|---|---|---|---|---|---|---|---|---|---|---|---|
| Scale | χ2 | | *scale** | *bkg* | $T_{shell}$ / Å | $R_{core}$ / Å | $\rho_{core}$ / Å$^{-2}$ | $\rho_{shell}$ / Å$^{-2}$ | $\rho_{solvent}$ / Å$^{-2}$ | $X_{shell}$ | $X_{core}$ | *bkg* | *α* | Scale |
| 1 | - | | 0.005 | $6.58 \cdot 10^{-4}$ | 7.66 | 11.88 | $8.40 \cdot 10^{-6}$ | $10.5 \cdot 10^{-6}$ | $9.40 \cdot 10^{-6}$ | 1 | Variable** | 0 | Variable | Variable** |

* the *scale* in the *Em* is set equal to the sample's volume fraction.

| **Full model** | | | **Core-Shell Bicelle (*Mm*)** | | | | | | | | | | **Power Law** | | |
|---|---|---|---|---|---|---|---|---|---|---|---|---|---|---|---|
| Scale | $\chi^2$ | | *scale** | *bkg* | Radius / Å | $T_{face}$ / Å | *L* / Å | $T_{rim}$ / Å | $\rho_{core}$ / Å$^{-2}$ | $\rho_{face}$ / Å$^{-2}$ | $\rho_{solvent}$ | $\rho_{rim}$ | *bkg* | *α* | Scale |
| 1 | - | | 0.005 | $4.10^{-4}$ | 500 | Variable** | Variable** | 0 | $8.40 \cdot 10^{-6}$ | $11.6 \cdot 10^{-6}$ | $9.40 \cdot 10^{-6}$ | / | 0 | Variable** | Variable** |

* the *scale* in the *Em* is set equal to the sample's volume fraction.

** the full set of variable parameters are reported as a function of pH in **Erreur ! Source du renvoi introuvable.** in th emain text

## Sample : RhaC10

Fixed parameters

| | **Core-Shell Ellipsoid (*Em*)** | | | | | **Power Law** | | |
|---|---|---|---|---|---|---|---|---|
| pH | *bkg* | *scale** | $\rho_{core}$ / Å$^{-2}$ | $\rho_{shell}$ / Å$^{-2}$ | $\rho_{solvent}$ / Å$^{-2}$ | *bkg* | *α* | Scale |
| 3.97 | $4.4 \cdot 10^{-4}$ | 0.025 | $8.40 \cdot 10^{-6}$ | $9.9 \cdot 10^{-6}$ | $9.40 \cdot 10^{-6}$ | 0 | 3.3 | $1.5 \cdot 10^{-8}$ |
| 5.11 | $4.4 \cdot 10^{-4}$ | 0.025 | $8.40 \cdot 10^{-6}$ | $9.9 \cdot 10^{-6}$ | $9.40 \cdot 10^{-6}$ | 0 | 3.3 | $1.5 \cdot 10^{-8}$ |
| 6.02 | $5.10^{-3}$ | 0.025 | $8.40 \cdot 10^{-6}$ | $9.4 \cdot 10^{-6}$ | $9.40 \cdot 10^{-6}$ | 0 | 3.3 | $1.5 \cdot 10^{-8}$ |

* the *scale* in the *Em* is set equal to the sample's volume fraction.

Variable parameters

| pH | Model | $R_{core}$ / Å | $T_{shell}$ / Å | $X_{core}$ | $X_{shell}$ |
|---|---|---|---|---|---|
| 3.97 | 1 | 8.6 | 10.6 | 3.1 | 1 |
| 3.97 | 1 | 8.3 | 11.4 | 3.9 | 0.5 |
| 3.97 | 1 | 8.0 | 11.9 | 4.7 | 0.1 |
| 5.11 | 1 | 8.3 | 9.5 | 2.5 | 1.1 |
| 6.02 | 1 | 6.1 | 3.4 | 9.1 | 1.1 |

**Table S 3 – Higher limits of the direct and derived structural parameters obtained by fitting the SAXS profiles of RL solutions and given in Table 1 and Table 4. $R_{core}$, $T_{shell}$ and $X_{core}$ contain a +10% error of the data given in Table 1. All other derived parameters are calculated with equations Eq. 4 - Eq. 10 and Eq. 14 – Eq. 16 for *PP*.**

| | **RhaRhaC10C10** | | **RhaC10C10** | **RhaC10** | | |
|---|---|---|---|---|---|---|
| pH | 6.05 | 7.5 | >7.5 | 3.97 | 5.11 | 6.02 |
| $R_{core}$ / Å | 9.1 | 8.8 | 13.1 | 9.5 | 9.1 | 6.7 |
| $T_{shell}$ / Å | 11.9 | 9.0 | 8.5 | 11.7 | 10.5 | 3.7 |
| $X_{core}$ | 3.4 | 2.8 | 2.8 | 3.4 | 2.8 | 10.0 |
| a ($R_{core}$ + $T_{shell}$) / Å | 21.0 | 17.8 | 21.6 | 21.1 | 19.6 | 10.5 |
| b ($R_{core}$ + $T_{shell}$) / Å | 21.0 | 17.8 | 21.6 | 21.1 | 19.6 | 10.5 |
| c ($R_{core}$ $X_{core}$ + $T_{shell}$) / Å | 43.0 | 33.2 | 44.5 | 43.9 | 35.6 | 70.9 |
| $V_{core}$ / Å$^3$ | 10865 | 7846 | 25824 | 12086 | 8762 | 12661 |
| $V_{shell}$ ($V_{ellips}$ - $V_{core}$) / Å$^3$ | 68627 | 36320 | 60715 | 69931 | 48310 | 19757 |
| $S_{core}$ / Å$^2$ | 2893 | 2203 | 4875 | 3106 | 2372 | 4432 |
| *Area/RL* / Å$^2$ | 129 | 136 | 92 | 62 | 66 | 85 |
| $n_{H2O/Rha(8)}$ | 2.0 | 1.8 | 0.6 | 0.8 | 1.0 | 0.1 |
| $n_{H2O/Rha(6)}$ | 1.3 | 1.2 | 0.4 | 0.5 | 0.7 | 0.0 |
| $N_{agg(8)}$ | 22 | 16 | 53 | 50 | 36 | 52 |
| $N_{agg(6)}$ | 29 | 21 | 68 | 64 | 46 | 67 |
| $PP_{(8)}$ | 0.41 | 0.40 | 0.40 | 0.41 | 0.40 | 0.43 |
| $PP_{(6)}$ | 0.32 | 0.31 | 0.31 | 0.32 | 0.31 | 0.33 |

**Table S 4 - Lower limits of the direct and derived structural parameters obtained by fitting the SAXS profiles of RL solutions and given in Table 1 and Table 4. $R_{core}$, $T_{shell}$ and $X_{core}$ contain a -10% error of the data given in Table 1. All other derived parameters are calculated with equations Eq. 4 - Eq. 10 and Eq. 14 – Eq. 16 for *PP*.**

| | **RhaRhaC10C10** | | **RhaC10C10** | **RhaC10** | | |
|---|---|---|---|---|---|---|
| pH | 6.05 | 7.5 | >7.5 | 3.97 | 5.11 | 6.02 |
| $R_{core}$ / Å | 7.5 | 7.2 | 10.7 | 7.7 | 7.5 | 5.5 |
| $T_{shell}$ / Å | 9.7 | 7.4 | 6.9 | 9.5 | 8.6 | 3.1 |
| $X_{core}$ | 2.8 | 2.3 | 2.3 | 2.8 | 2.3 | 8.2 |
| a ($R_{core}$ + $T_{shell}$) / Å | 17.2 | 14.6 | 17.6 | 17.3 | 16.0 | 8.6 |
| b ($R_{core}$ + $T_{shell}$) / Å | 17.2 | 14.6 | 17.6 | 17.3 | 16.0 | 8.6 |
| c ($R_{core}$ $X_{core}$ + $T_{shell}$) / Å | 30.6 | 23.6 | 31.0 | 31.1 | 25.4 | 48.0 |
| $V_{core}$ / Å$^3$ | 4869 | 3516 | 11572 | 5416 | 3927 | 5674 |
| $V_{shell}$ ($V_{ellips}$ - $V_{core}$) / Å$^3$ | 32940 | 17470 | 28849 | 33506 | 23319 | 9024 |
| $S_{core}$ / Å$^2$ | 1609 | 1232 | 2726 | 1727 | 1326 | 2435 |
| *Area/RL* / Å$^2$ | 160 | 170 | 114 | 77 | 82 | 104 |
| $n_{H2O/Rha(8)}$ | 4.5 | 4.1 | 1.3 | 1.8 | 2.4 | 0.2 |
| $n_{H2O/Rha(6)}$ | 3.1 | 2.9 | 0.9 | 1.3 | 1.7 | 0.0 |
| $N_{agg(8)}$ | 10 | 7 | 24 | 22 | 16 | 23 |
| $N_{agg(6)}$ | 13 | 9 | 31 | 29 | 21 | 30 |
| $PP_{(8)}$ | 0.41 | 0.40 | 0.40 | 0.41 | 0.40 | 0.42 |
| $PP_{(6)}$ | 0.32 | 0.31 | 0.31 | 0.32 | 0.31 | 0.33 |

**Table S 5 – Structural molecular parameters employed to calculate molecular volumes**

**Rhamnose, Rh, $C_6H_{12}O_5$**

| | | |
|---|---|---|
| Molar volume | 113.89 | $cm^3$/mol |
| Molar volume | $1.1389 \cdot 10^{26}$ | $Å^3$/mol |
| Avogadro | $6.02 \cdot 10^{26}$ | molec/mol |
| Volume/Rh | 189 | $Å^3$/Rh |

**Formic acid, FA, $CH_2O_2$**

| | | |
|---|---|---|
| Molar volume | 37.8 | $cm^3$/mol |
| Molar volume | $3.78 \cdot 10^{25}$ | $Å^3$/mol |
| Volume/FA | 63 | $Å^3$/FA |

**Water, W, $H_2O$**

| | | |
|---|---|---|
| Molar volume | 18 | $cm^3$/mol |
| Molar volume | $1.8 \cdot 10^{25}$ | $Å^3$/mol |
| Volume/W | 30 | $Å^3$/W |
| Volume $(CH_2)_8CH_3$ | 243 | $Å^3$ |
| Volume $(CH_2)_6CH_3$ | 189 | $Å^4$ |
| Volume $(CH_2)_2$ | 54 | $Å^3$ |

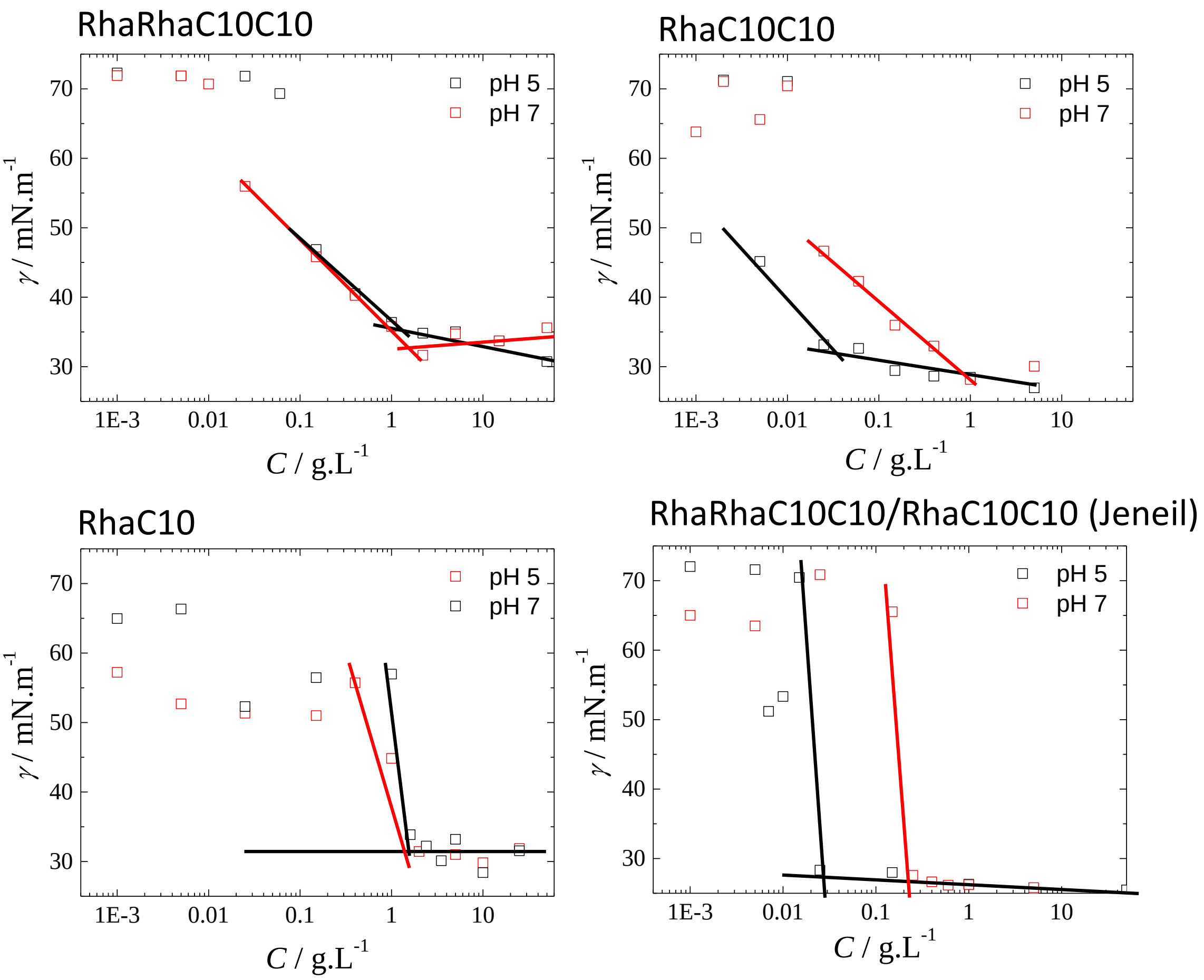

**Figure S 1 – Surface tension experiments performed on RL solutions to determine their critical micelle concentration (cmc).**

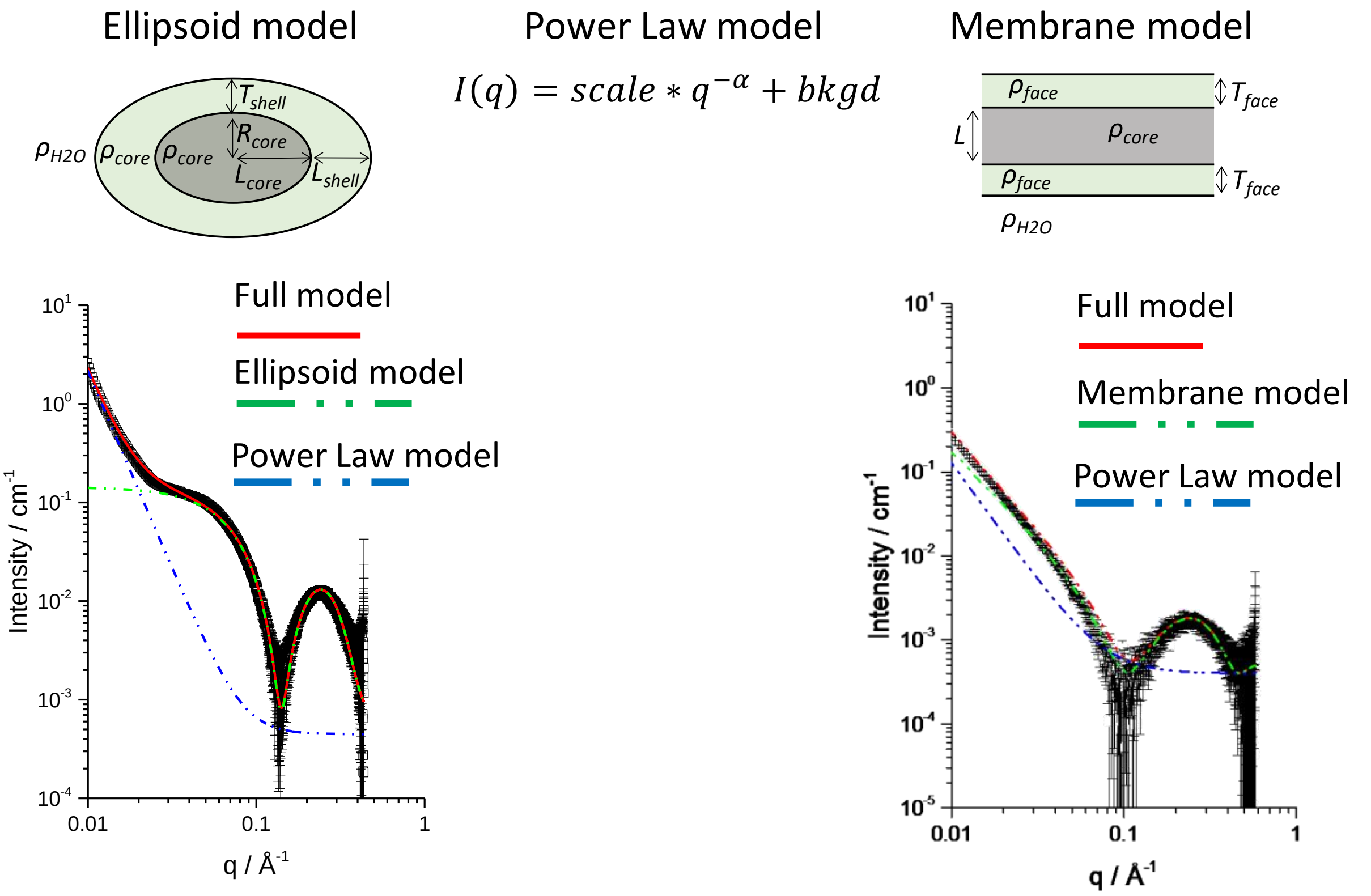

**Figure S 2 – Typical contributions of the individual *Em*, *Mm* and *Power law* models to the fit of the entire SAXS profile.**

The solubility and stability of C10C10 dependant on the pH value was investigated. For this purpose, the solubility was first analyzed by means of turbidity measurements after adjustment of the pH (Figure S 3a). It was found that C10C10 showed a distinct turbidity of the solutions already below pH 8. The data obtained by turbidity measurements were supplemented by HPLC analysis, which allowed the C10C10 concentration to be monitored over time. Figure S 3b confirms the decrease in solubility in acidic environments over time.

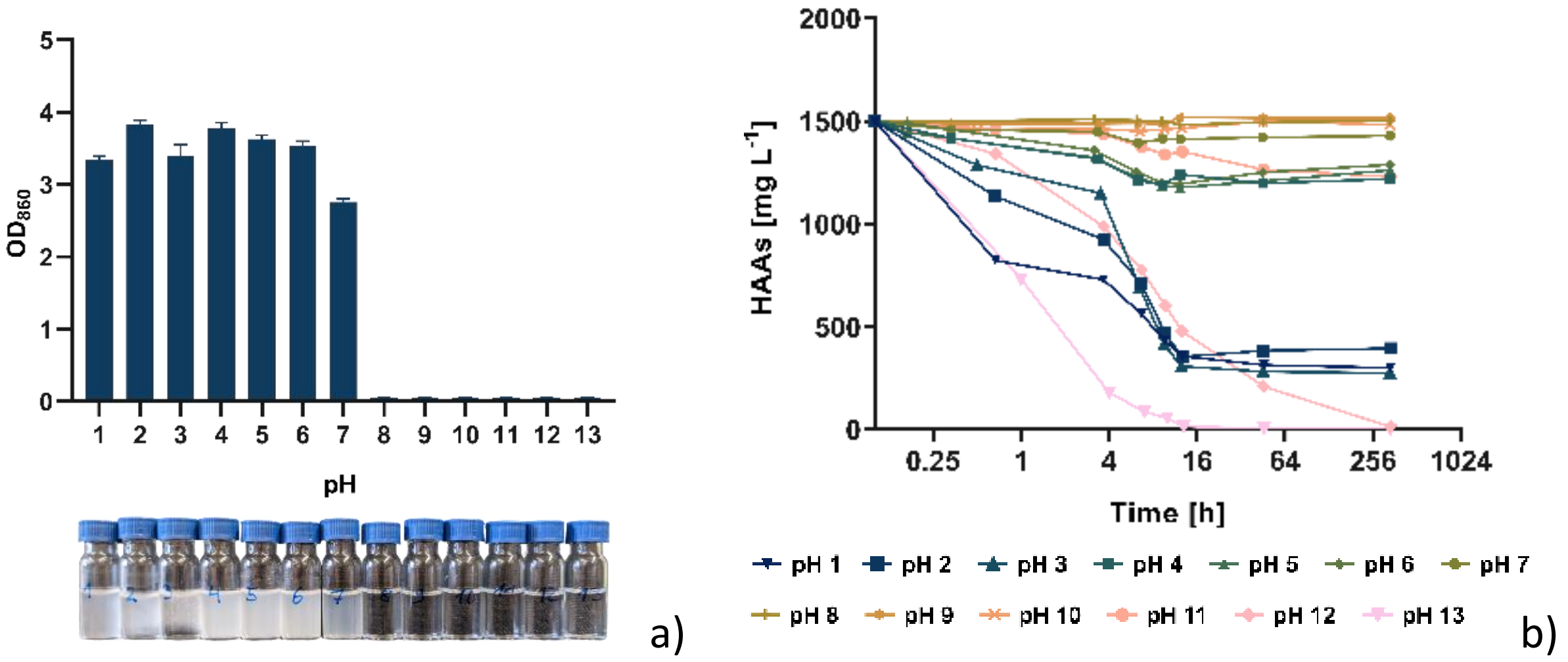

**Figure S 3 – Turbidity of C10C10 solutions at different pH values determined via optical density at a wavelength of 860 nm immediately after sample preparation. The error bars indicate the deviation from the mean of three analytical replicates.**

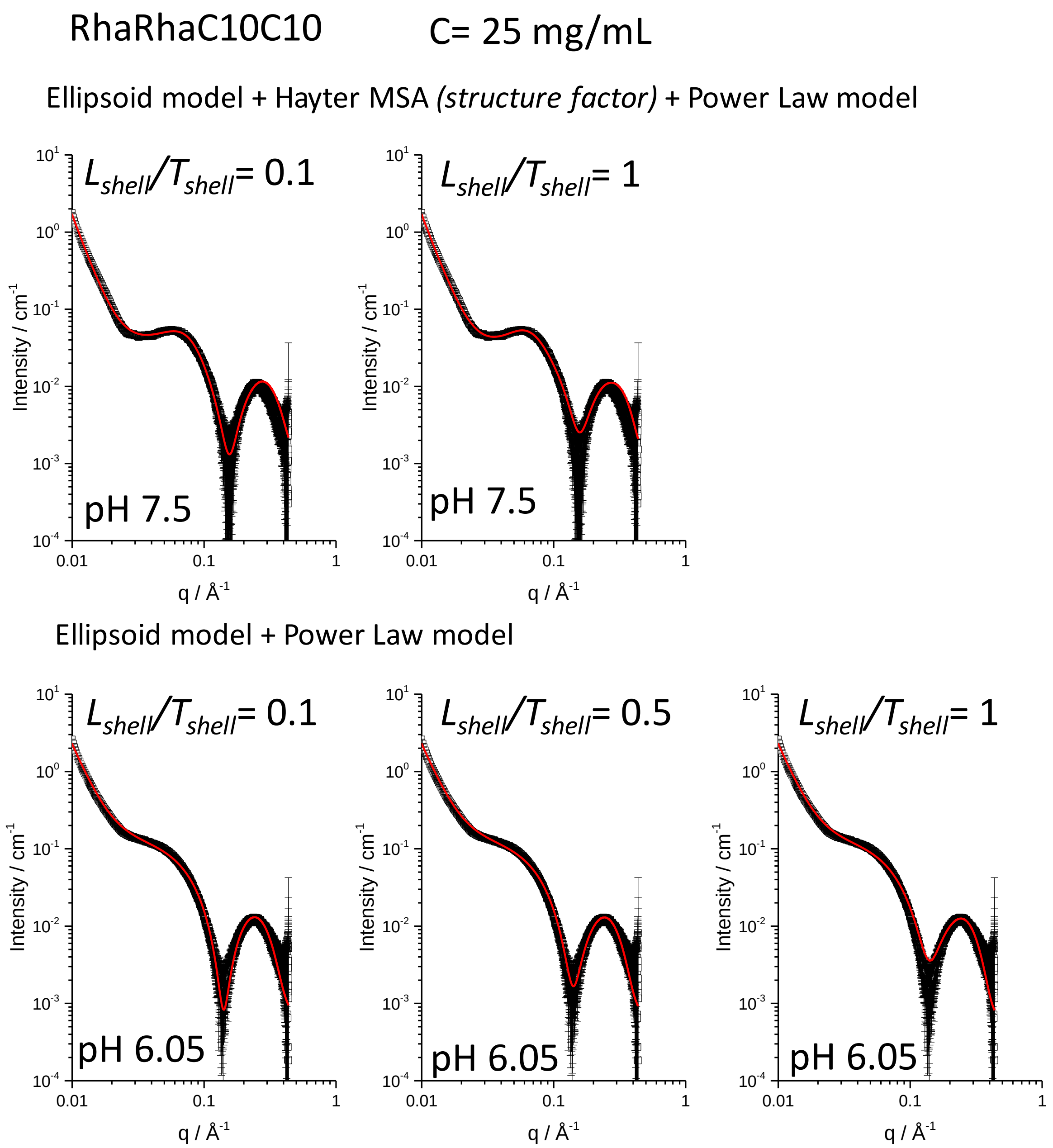

**Figure S 4 – Fits of the SAXS profiles of RhaRhaC10C10 solutions at pH 7.5 and pH 6.05. Different values of the shell aspect ratio (*$L_{shell}$/$T_{shell}$*, please refer to *Em* model in Figure 2 and parameters in Table 1) are used in the fitting process.**

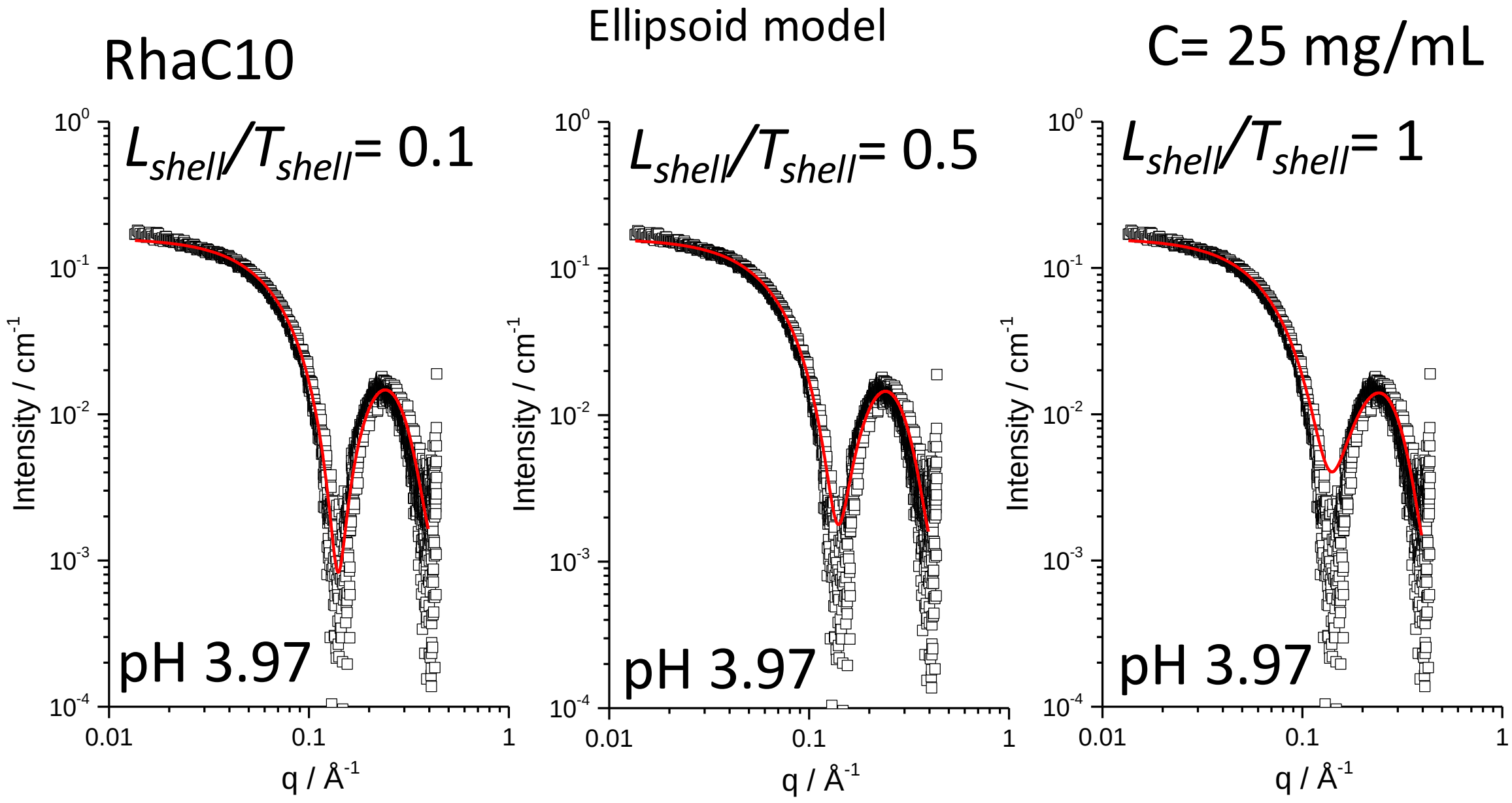

**Figure S 5 – Fits of the SAXS profiles of the RhaC10 solution at pH 3.97. Different values of the shell aspect ratio ($L_{shell}/T_S$, please refer to *Em* model in Figure 2 and parameters in Table 1) are used in the fitting process.**

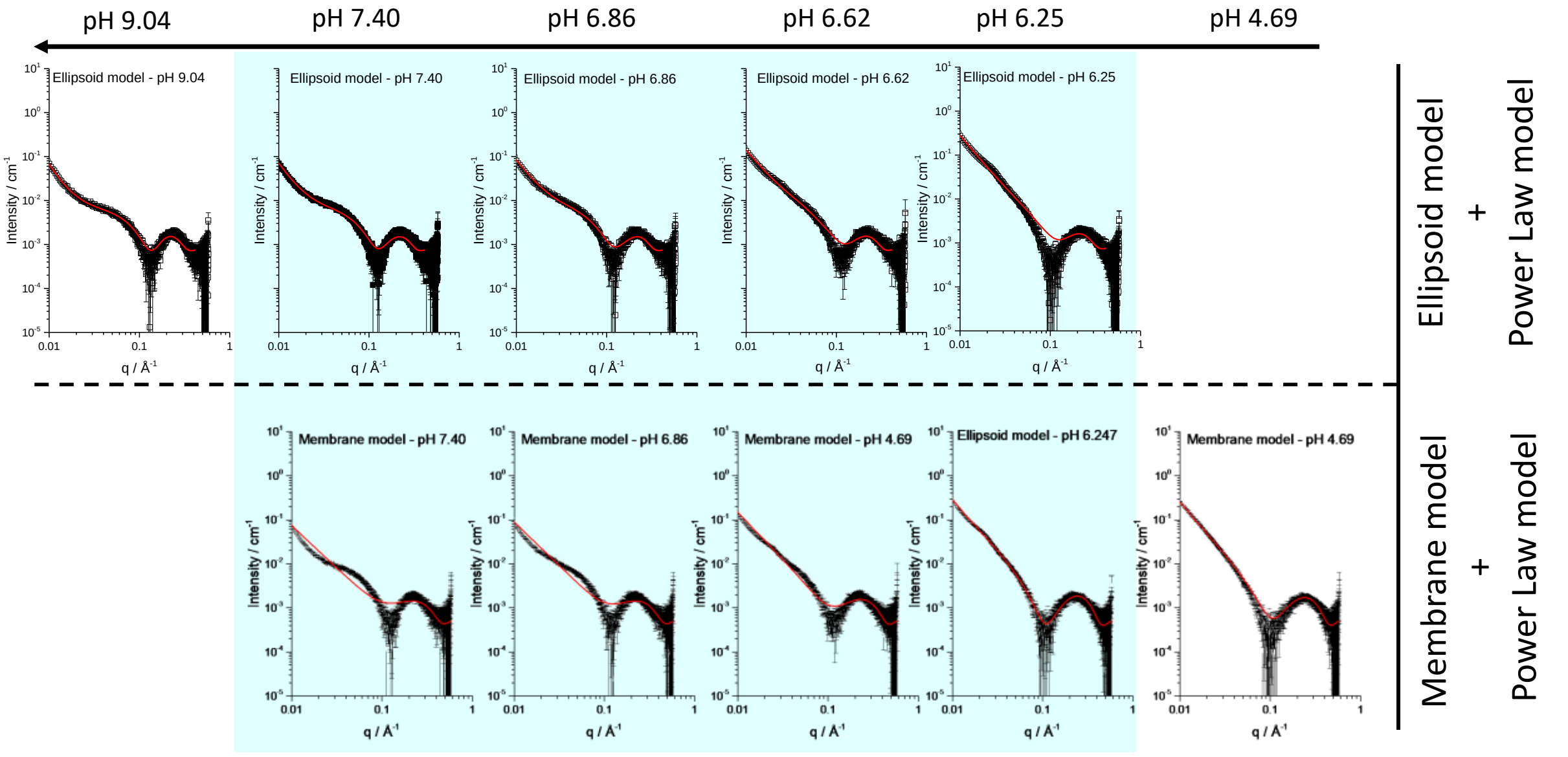

**Figure S 6 – Series of SAXS profiles of RhaC10C10 solutions extracted from the pH-resolved *in situ* experiment in Figure 7. In the top line, fitting is performed using the *Em* + Power law models. In the bottom line, fitting is performed using the *Mm* + Power law models. The list of fitting parameters is given in Table S2. The *Em* + Power law model fits well the SAXS profiles at basic pH and poorly at acidic pH, while the *Mm* + power law model fits well the SAXS data at acidic pH and poorly at basic pH.**